**Molecular Hyperpolarisability as a Screening Descriptor for Second-Order Nonlinear Optics in Ferroelectric Nematic Liquid Crystals**

Charles Parton-Barr[1], Nerea Sebastian[2], Richard J. Mandle [1,3] *

[1]School of Physics and Astronomy, University of Leeds, Leeds, UK, LS2 9JT
[2] Jožef Stefan Institute, Ljubljana, Slovenia
[3] School of Chemistry, University of Leeds, Leeds, UK, LS2 9JT

**Abstract**

Ferroelectric nematic liquid crystals combine fluidity with macroscopic polar order. Although the archetypal $N_F$ material RM734 exhibits large nonlinear optical coefficients, most known ferroelectric nematics were designed without consideration of optical nonlinearity. Here, we assess whether electronic-structure calculations can be used to identify promising nonlinear optical candidates within known polar liquid-crystal materials. Frequency-dependent molecular hyperpolarisability tensors were calculated using a range of DFT methods and basis sets, then converted to macroscopic d-coefficients using an oriented-gas model with empirical $\langle P_1 \rangle$ and $\langle P_3 \rangle$ values. Calculated values were benchmarked against available experimental $d_{33}$, $d_{15}$, and $d_{13}/d_{31}$ coefficients for representative materials. We find that absolute values depend strongly on method and bulk-parameter assumptions, whereas relative trends are more useful for screening. Explicit conformer averaging does not consistently improve agreement with experiment. Applying the best-performing single-conformer protocols to a broader polar liquid-crystal dataset identifies candidate materials with enhanced predicted nonlinear optical response and reveals simple design rules based on donor-acceptor asymmetry, conjugation length, and linker choice.

**Introduction**

The development of new material systems with tailored nonlinear optical responses is central to advancing a wide range of optical phenomena that enable modern photonics. Soft matter platforms that can support non-centrosymmetric order while maintaining structural flexibility and tunability under external stimuli are of great interest, offering alternative routes to conventional crystalline or polymer-based nonlinear optical media.

Second-order nonlinear optical (NLO) phenomena arise when the polarisation of a material responds quadratically to an applied electric field. The macroscopic polarisation of a material can be written in the following functional form

$$P_i = \varepsilon_0 \left( \chi_{ij}^{(1)} E_j + \chi_{ijk}^{(2)} E_j E_k + \chi_{ijkl}^{(3)} E_j E_k E_l + \cdots \right) \quad (1)$$

Here $\chi_{ij}^{(1)}$, $\chi_{ijk}^{(2)}$ and $\chi_{ijkl}^{(3)}$ are the linear, second-order and third-order electric susceptibilities, respectively. The second-order term is responsible for processes such as second harmonic generation for frequency doubling in photonic applications or spontaneous parametric down-conversion for photon-pair generation in quantum optics, and is commonly expressed using the contracted nonlinear optical coefficients,

$$d_{il} = \frac{1}{2} \chi_{\mathrm{ijk}}^{(2)} \quad (2)$$

Here, the pair of indices $(jk)$ is replaced by the contracted index $(l)$. Permutation symmetries allow us to replace the $j$ and $k$ indices with $l$, giving a $3 \times 6$ matrix of coefficients, $d_{il}$. A non-zero $\chi^2$ requires broken inversion symmetry; in a centrosymmetric medium the second-order susceptibility is zero. In conventional nematics the director is *apolar* such that $+n \equiv -n$ and the phase is therefore centrosymmetric on average. Similarly, antiferroelectric smectic phases ($SmA_{AF}$ [1] and $SmC_{AF}$ [2]) are also centrosymmetric on average. Recently discovered ferroelectric nematic ($N_F$) liquid crystals [3-8] and analogous polar smectic phases such as $SmA_F$ [9, 10] and $SmC_P$ [11, 12] are therefore attractive candidates for second-order NLO materials because they provide a route to non-centrosymmetric organisation in a soft, fluid medium. These recently discovered ferroelectric phases have far larger spontaneous polarisation than the SmC* ferroelectrics of long ago.

Ferroelectric liquid crystals can be reoriented with small applied electric fields [13, 14] unlike competing solid-state materials where the NLO response is related to specific crystal axes. Moreover, ferroelectric liquid crystals benefit from polar ordering emerging *spontaneously*, whereas polymeric (or glassy) systems can require poling with large external fields. Their potential motivated a range of demonstrations, including their use for reconfigurable Pancharatnam-Berry nonlinear diffractive elements [15], for nonlinear optical structures with geometric phase encoding [16], GHz-rate modulators in silicon photonic platforms [17] or as tuneable sources of entangled photon pairs [18, 19].

RM734 has been used as a benchmark $N_F$ material in a substantial number of studies. While RM734 is reported to have remarkably large nonlinear optical coefficients in the $N_F$ phase [20, 21], it (and most other $N_F$ materials, so far as we are aware) was not designed as an NLO chromophore. The structural features compatible with polar liquid-crystal organisation that also maximise molecular hyperpolarisability ($\beta$) – and hence the macroscopic coefficients $d_{31}$, $d_{33}$, $d_{15}$ – are poorly established. This limits the rational discovery of new NLO-optimised $N_F$ materials.

A natural way to address this limitation is to use electronic structure calculations as a screening tool. The molecular first hyperpolarisability tensor ($\beta_{ijk}$) can be calculated using quantum-chemical methods at frequencies consistent with experimental measurement. This can then be related to the macroscopic $d_{il}$ coefficients through a simple orientational averaging model. In this approach the molecular response provides the microscopic origin of the bulk second-order susceptibility, while the liquid-crystalline phase provides the polar and orientational order required for a non-zero macroscopic response. It is unknown, however, whether a single molecular geometry can adequately describe $\beta_{ijk}$, or if population weighted conformers are required. By identifying electronic structure methods capable of reproducing experimental values for widely known materials (RM734, [3] DIO, [5] C1; [22] see ESI for structures), we can then leverage these methods to identify promising NLO candidates from the known pool of $N_F$ materials and devise strategies for increasing $\beta$ and the resulting $d_{31}$, $d_{15}$, $d_{33}$ coefficients.

## Methods

### Electronic Structure Calculations

Initial geometries were optimised with the GFN2-xTB semi-empirical tight-binding method [23]; conformers were generated with the GOAT algorithm in ORCA 6.1 at this same level [24]. Conformers were retained to capture 80% of the cumulative Boltzmann population for RM734 (7 conformers) and DIO (30 conformers).

The frequency-dependent hyperpolarisability tensor ($\beta_{ijk}$) was computed at incident wavelengths of 800 nm using a range of methods in the Gaussian G16 rev.c01 software package [25]. Calculations were performed either on a single geometry optimised at the B3LYP/6-311+G(d,p) level of DFT, or on the conformer ensemble generated as above. For conformer-ensemble calculations, the hyperpolarisability tensor of each conformer was weighted by its Boltzmann population, calculated using the SCF energy obtained at the same level of theory.

Methods were chosen from the pool of methods available in Gaussian 16 that support frequency-dependent hyperpolarisability calculations. We utilise Hartree-Fock as a low-level reference (HF [26]), hybrid GGA DFT (B3LYP [27], BHandHLYP [28, 29]), hybrid mGGA DFT (M06 [30], M06-2X [30], M06-HF [31, 32], TPSSH [33, 34]), and range-separated hybrid DFT (CAM-B3LYP [35], wB97X-D [36], LC-BLYP [37]). The chosen methods span the rungs of "Jacob's ladder" [38] from the Hartree-Fock method through to range-separated hybrids. These methods were paired with 6-31+G(d) [39-42], 6-311++G(d,p) [43-45], aug-cc-pVDZ and aug-cc-pVTZ [46, 47] basis sets. The use of large basis sets with diffuse and polarisable functions is especially important to accurately describe the electronic response to an applied field.

**Orientational Averaging of the Hyperpolarisability Tensor**

The thermally averaged hyperpolarisability components in the laboratory frame were evaluated using expressions reported in Refs [21, 48].

$$\langle\beta_{33}\rangle = \frac{1}{5}\left[(\langle P_1\rangle - \langle P_3\rangle)\left(2\beta_{yyx} + 2\beta_{zzx} + \beta_{xyy} + \beta_{xzz}\right) + (3\langle P_1\rangle + 2\langle P_3\rangle)\beta_{xxx}\right] \quad (3.1)$$

$$\langle\beta_{15}\rangle = \frac{1}{10}\left[\langle P_3\rangle\left(2\beta_{yyx} + 2\beta_{zzx} + \beta_{xyy} + \beta_{xzz} - 2\beta_{xxx}\right) + \langle P_1\rangle\left(3\beta_{yyx} + 3\beta_{zzx} - \beta_{xyy} - \beta_{xzz} + 2\beta_{xxx}\right)\right] \quad (3.2)$$

$$\langle\beta_{31}\rangle = \frac{1}{10}\left[\langle P_3\rangle\left(2\beta_{yyx} + 2\beta_{zzx} + \beta_{xyy} + \beta_{xzz} - 2\beta_{xxx}\right) - 2\langle P_1\rangle\left(\beta_{yyx} + \beta_{zzx} - 2\beta_{xyy} - 2\beta_{xzz} - \beta_{xxx}\right)\right] \quad (3.3)$$

These relations provide the orientationally averaged laboratory-frame tensor $\langle\beta_{ijk}\rangle$ in terms of the molecular-frame hyperpolarisability components $\beta_{lmn}$ (with molecular geometries oriented such that the long axis is along the x-axis) and the orientational order parameters $\langle P_1\rangle$ and $\langle P_3\rangle$. The components $\beta_{lmn}$ are obtained from electronic structure calculations at the different Gaussian levels of theory. The molecular tensor is then transformed into the laboratory frame defined such that the nematic director coincides with the x-axis. Assuming the uniaxial symmetry of the ferroelectric nematic phase, the resulting laboratory-frame tensor is averaged over the azimuthal orientations, yielding an expression $\langle\beta_{IJK}^{lab}\rangle$ in which the laboratory-frame orientational dependence is only on the polar angle $\theta$.

The thermally averaged components are then obtained by orientationally averaging over the molecular ensemble using an orientational distribution function (ODF) based on a  expansion in terms of Legendre polynomials

$$f(cos\theta) = \sum_{l=0}^{\infty} \frac{2l+1}{2} \langle P_l \rangle P_l(\cos\theta) \tag{4}$$

Here, $\langle P_l \rangle$ are orientational order parameters and $P_l(\cos\theta)$ are the Legendre polynomials, and

$$\langle \beta_{ijk} \rangle = \int_0^{\pi} \langle \beta_{ijk}^{lab} \rangle f(cos\theta) \sin(\theta)\, d\theta \tag{5}$$

In a uniaxial polar phase, the orientational averaging of the third-rank hyperpolarisability reduces to linear combinations of molecular tensor elements weighted by the orientational order parameters $\langle P_1 \rangle$ and $\langle P_3 \rangle$ given in expressions 3.1-3.3.

The values $\langle P_1 \rangle$ and $\langle P_3 \rangle$ can be experimentally determined as proposed in reference [20], considering that the molecular orientation distribution is governed by a phenomenological mean-field orientational potential, in which the apolar nematic mean-field term is accompanied by an additional effective potential representing polar ordering. $\langle P_1 \rangle$ is obtained from polarization measurements, while $\langle P_2 \rangle$ can be connected to the birefringence ratio between the values in the polar phase and the extrapolated value from the non-polar phase. These two conditions allow us to determine the parameters of the orientational potential, which define the molecular orientational distribution. The third-rank order parameter $\langle P_3 \rangle$ is then calculated from this distribution by evaluating the corresponding orientational average. For RM734, DIO and C1 values thus obtained are taken from reference [21]. Considering the similar values reported for RM734 and DIO, both in the $N_F$ phase, representative values of $\langle P_1 \rangle$ = (0.82, 0.78, 0.69) and $\langle P_3 \rangle$ = (0.48, 0.59, 0.62) were adopted for RM734, DIO and C1, respectively. A systematic experimental determination of these parameters for all screened materials would be desirable but is beyond the scope of the present computational study, and thus we use RM734 values for reference.

**Oriented-gas model, local field and number density**

From the thermally averaged hyperpolarisability tensor we calculated the d tensor components through an oriented gas model [49].

$$d_{IJK} = \frac{N}{4\varepsilon_0} f^3 \langle \beta_{ijk} \rangle, \qquad N = \frac{N_A \rho}{M} \tag{6}$$

Here, $f$ is the field factor and $N$ is the number density calculated from material density [50], $\rho$, and molar mass, $M$. The factor of ¼ is necessary due to a combination of the Taylor series representation of the induced dipole used in Gaussian 16 and a degeneracy factor that arises in SHG processes when converting from microscopic to macroscopic representations [51]. An isotropic approximation was adopted for the local field corrections using extraordinary and ordinary refractive indices ($n_e = 1.8$, $n_o$=1.5). The refractive indices were used to calculate the Lorentz factors, $f = \frac{\bar{n}^2+2}{3}$ use the Vuks effective refractive index $\bar{n}^2 = \frac{n_e^2+2n_o^2}{3}$. This treatment is justified and follows the methodology described in ref. [21]. When applying this method to the broader polar liquid-crystal dataset, reference values for RM734 were considered. $n_o = 1.5$ provides a well-justified approximation across all systems. Although $n_e$ may vary more substantially, taking $n_e = 1.8$ as a representative upper bound leads to changes in the Lorentz factors that remain confined to the typical experimental spread (1.65–1.8), corresponding to a variation of about 11%, indicating a relatively small effect in this approximation on the macroscopic response compared to molecular contributions. The refractive indices choices are justified by literature values [21, 52]. All calculations used a density of 1.3 g cm$^{-3}$, measured for the $N_F$ material M5 [50].

## Dataset

A curated dataset of all reported polar liquid crystals ($N_F$, $SmA_F$, $SmC_P$ and variations thereof) was built from the literature; detailed in ref. [53], data taken from refs [1, 3-5, 22, 54-107]. We evaluate each material using the benchmarked single-conformer workflow developed herein. Materials were ranked by their calculated NLO coefficients, and the resulting values were interpreted alongside molecular structure, allowing us to identify chemical motifs associated with large predicted nonlinear optical response.

## Results and Discussion

Using the workflow outlined above, we set out to identify a practical electronic structure method for calculation of NLO coefficients for polar liquid crystalline materials. Experimentally measured NLO coefficients exist for RM734, DIO, and C1 and these were used to benchmark different calculation types, and to explore the benefit (or lack of) offered by considering multiple conformations.

## Calculations on Global Minimum Energy Geometries

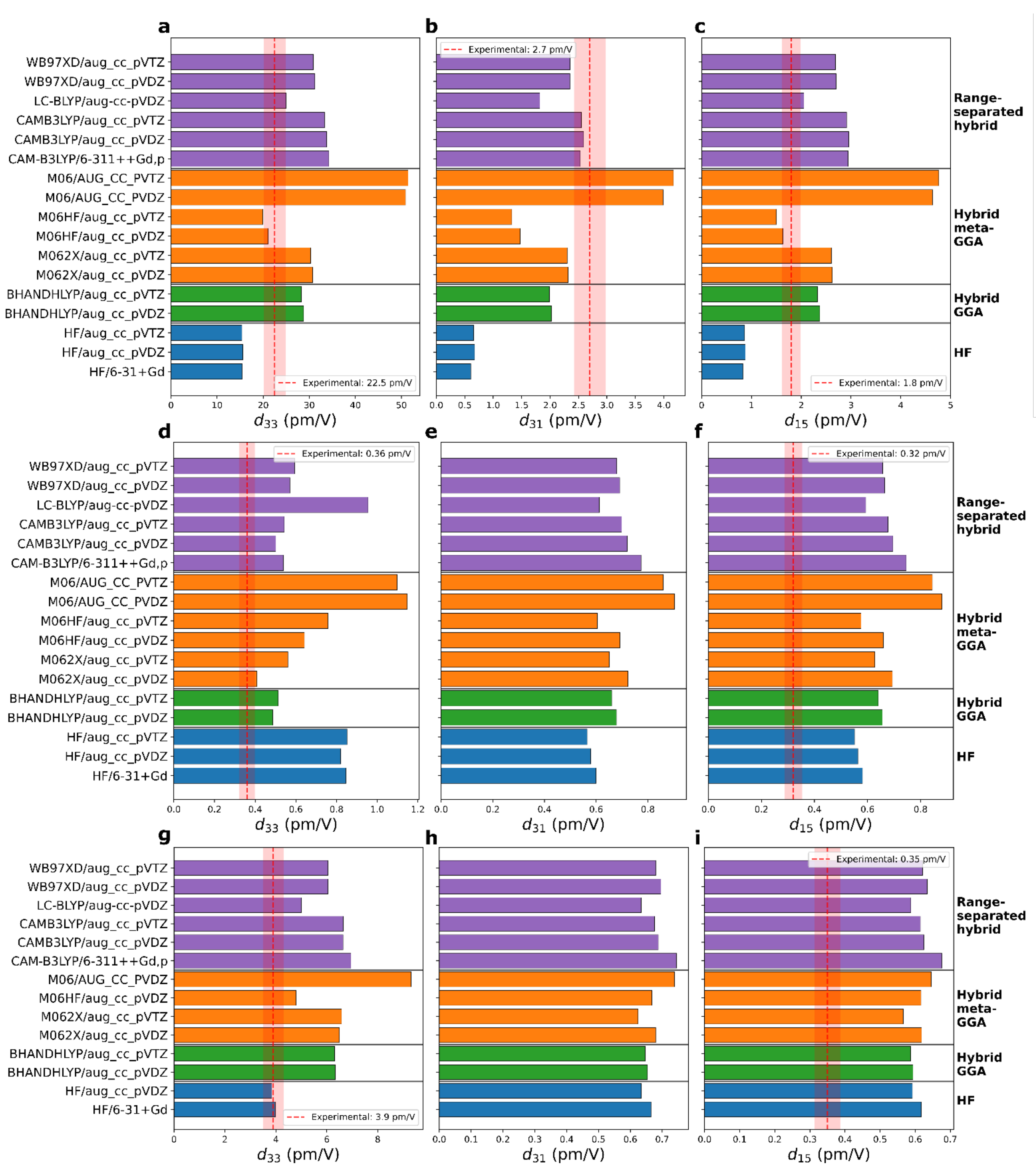

**Figure 1:** Calculations of d coefficients for ground state geometries of RM734 (a-c), DIO (d-f), and C1 (g-i). The calculated d coefficients are grouped into method families which are given on the right-hand side of the figure. Comparisons are made to the experimental values (dashed red line, shaded area

corresponding to an estimated 10% standard error in this value) wherever possible.

We first perform calculations using a single geometry; we identify the global minimum with the GOAT algorithm at the GFN2-xTB level and reoptimise this at the B3LYP/6-311+G(d,p) level of DFT. Calculated d coefficients for RM734, DIO, and C1 using the methodology described above are given in Figure 1 and compared to experimental values wherever possible. In the calculations of C1 some calculations are not presented due to severe overestimation of the d coefficients or a failure of some C1 aug-cc-pVTZ calculations to converge. Calculated d coefficients for all materials are given in the ESI.

The calculated d coefficients reproduce the experimental ordering of the coefficient magnitudes reasonably well (Figure 1), suggesting that the underlying $\beta$ tensor components are being captured at a physically sensible scale. However, an examination of the results presents several issues.

The HF calculations underpredict the experimental value of $d_{33}$, $d_{31}$ and $d_{15}$ for RM734 irrespective of the basis set. For DIO, these same methods consistently overestimate the d coefficients and give among the poorest predictions of $d_{33}$. For C1, however, HF/aug-cc-pVDZ gives the most accurate $d_{33}$ calculation of any method, while the $d_{15}$ value obtained with this method is in line with those from other, more expensive methods.

The widely used hybrid GGA method B3LYP is omitted from Figure 1 for each molecule under study because of its universally poor performance here. For RM734, BHandHLYP overestimates $d_{33}$ and $d_{15}$ while underestimating $d_{31}$. This apparent difference between $d_{15}$ and $d_{31}$ should not be overinterpreted: across the tested methods the calculated values of these two coefficients are very similar, and for RM734 their separation is comparable to the experimental difference of just 0.9 $pm/V$. Similarly, BHandHLYP overpredicts the $d_{33}$ and $d_{15}$ coefficients for both DIO and C1, indicating a systematic tendency towards larger calculated coefficients for this functional.

The hybrid meta-GGA methods show a clear dependence on exact-exchange character. M06 gives the largest coefficients and substantially overestimates the experimental values, while M06-HF gives the smallest values and provides one of the better estimates of $d_{33}$ for RM734 and C1. M06-2X generally lies between these limits, reflecting its intermediate degree of HF exchange. For DIO, the ordering is less clean, particularly for $d_{15}$, but the same broad trend is retained.

The range-separated hybrids generally overpredict RM734's $d_{33}$ coefficient compared to most methods except for M06; LC-BLYP underpredicts compared to the rest of this family. The same qualitative behaviour is observed for $d_{31}$ and $d_{15}$, with the relative ordering of methods largely retained across the different d coefficients. For DIO and C1, all methods consistently overpredict $d_{33}$ and $d_{15}$.

Throughout Figure 1, the choice between aug-cc-pVDZ and aug-cc-pVTZ does not play a substantial role in the magnitude of the calculated d coefficients and, given the increased cost of the triple-zeta basis sets, this is rather welcome. These results demonstrate that the available electronic structure methods for a single low-energy conformer can reproduce the trends in experimental NLO coefficients, through an adequate calculation of the underlying $\beta_{ijk}$ tensor; however, the magnitudes of the calculated coefficients are ultimately dependent on the assumed parameters.

### Calculations on Conformer Ensembles

Each of the three molecules studied possesses several low-energy conformations, arising from rotation of esters, inter-ring torsions, and related degrees of freedom. We therefore considered whether an ensemble of low-energy conformers might reduce the discrepancy between experimental and calculated values. We note that calculations of hyperpolarisability tensors with large diffuse basis sets are already computationally demanding, and the use of an ensemble of conformers makes it necessary to run these calculations multiple times for different input geometries. As C1 has several hundred low-energy conformers we performed this analysis for RM734 and DIO only.

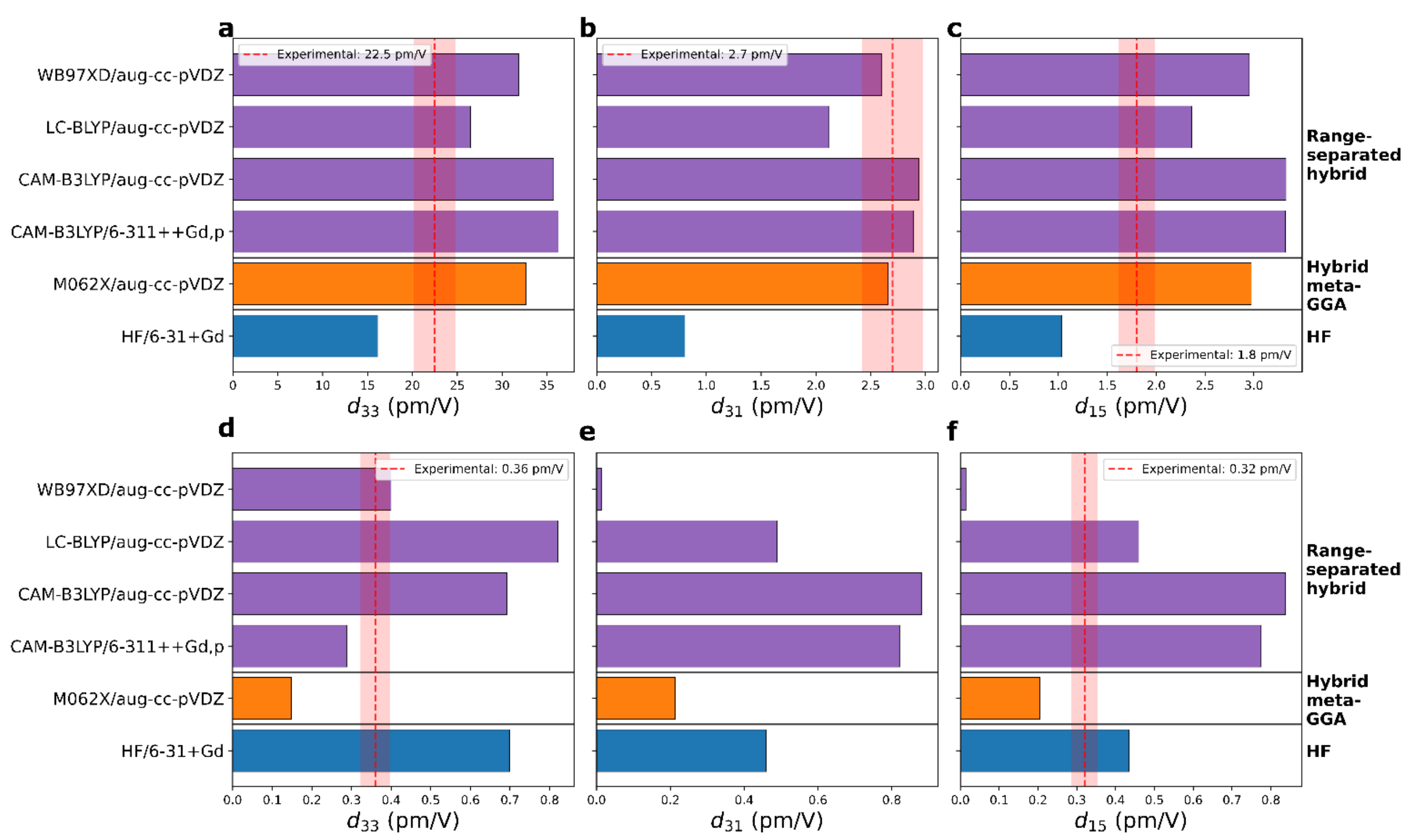

**Figure 2:** Calculations of d coefficients for conformer ensemble geometries of RM734 (a-c) and DIO (d-f)

Against our initial expectations, as shown in **Figure 2**, taking $\beta$ as a weighted average over the conformer ensemble does not significantly improve the correlation between experimental and calculated NLO coefficients for any tested functional/basis set combination. For RM734 the values are all broadly similar, with the conformer ensemble leading to a consistent increase in calculated $d_{15}$ and $d_{31}$ values. This may be due to the relatively inflexible nature of the molecule. DIO has more variation between the ground state and conformational predictions. For DIO, CAM-B3LYP/ 6-311++G(d,p) gives the closest value of $d_{33}$ when conformer averaging is included, although its prediction of $d_{15}$ remains poor. LC-BLYP, and HF all give reduced $d_{33}$ values, while wB97X-D is slightly increased. M06-2X gives a value less than half that of the single conformer (0.15 $pm/V$ versus 0.41 $pm/V$). A visual comparison between the results for the two geometry conditions is given in the ESI (Figure SI-1).

### Method Selection

The variance seen across materials and methods motivates further analysis of the ratios of the calculated d coefficients between molecules and how they compare to their experimental counterparts. This requires the assumption that the assumed parameters from equation 6 are similar enough to cancel, giving us the relationship

$$\frac{d_{exp1}}{d_{exp2}} \approx \frac{d_{calc1}}{d_{calc2}} \approx \frac{\langle \beta_{ijk1} \rangle}{\langle \beta_{ijk2} \rangle} \tag{7}$$

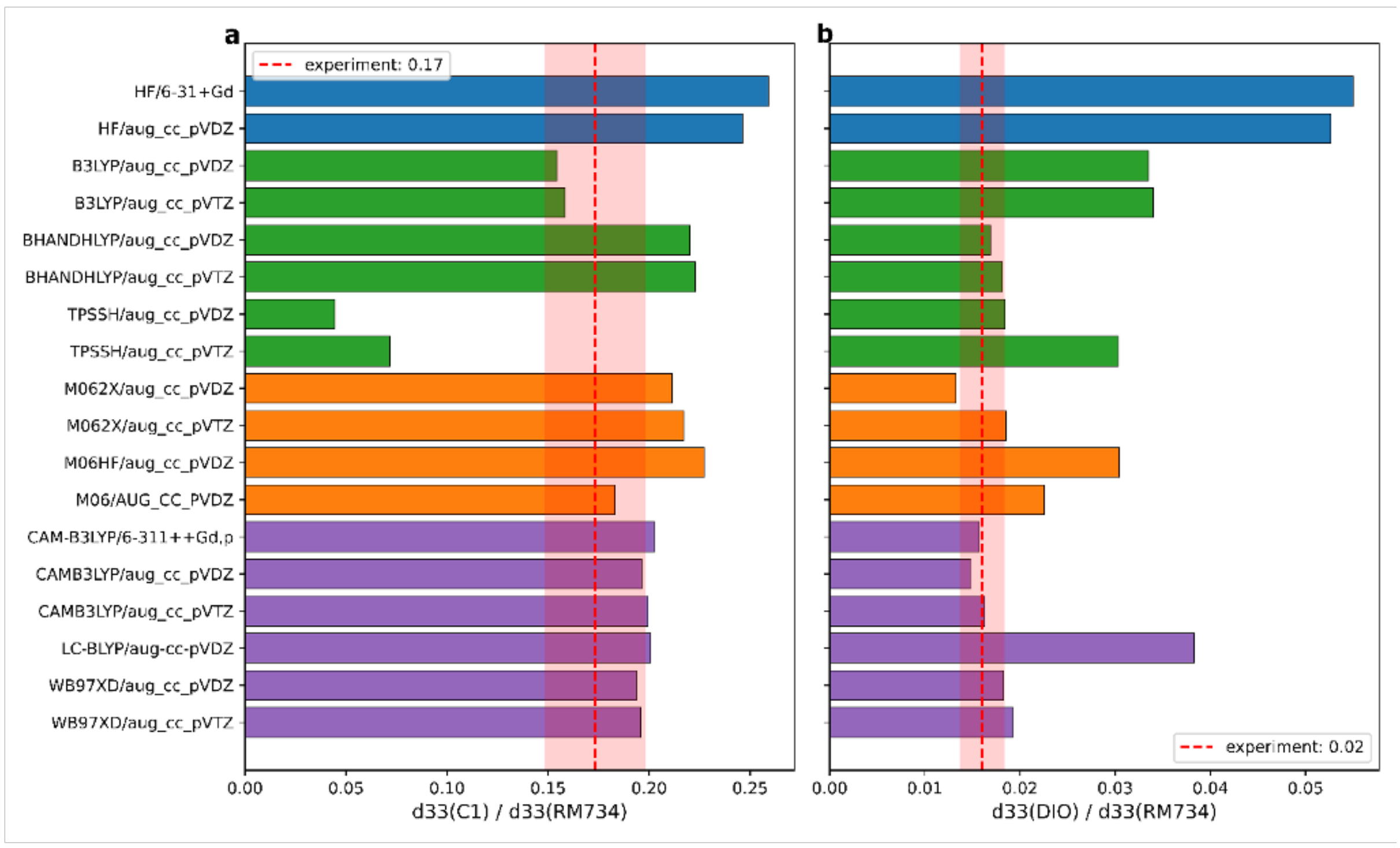

**Figure 3.** Performance of different functional/basis set combinations in predicting $d_{33}$ expressed as a ratio of the calculated $d_{33}$ for C1 or DIO and RM734. Both plots use data for single conformers only. The dashed red line corresponds to the experimental value of this ratio, and the shaded area corresponds to an estimated ~14% standard error in this value due to propagation).

We use RM734 as the reference molecule to calculate $d_{33}$ ratios for all the methods and basis sets with the previously stated assumed parameters (e.g. density, $\langle P_1 \rangle$ and so on). We selected $d_{33}$ for analysis here as this is the predominant coefficient in the studied materials and thus is the one of interest for second-order nonlinear optical processes. As shown in Figure 3, this gives a more easily interpretable sense of how different methods compare across molecules. Hartree-Fock gives values which correlate poorly against experiment. For RM734/C1 we can see that the reproduction of the experimental ratio is stable across the range-separated hybrids; however, for DIO/RM734 there is less stability with this range-separated family. Figure 3 also shows that over- or underprediction is not consistent across both molecular comparisons; a method that performs well for RM734/C1 does not necessarily perform well for DIO/RM734.

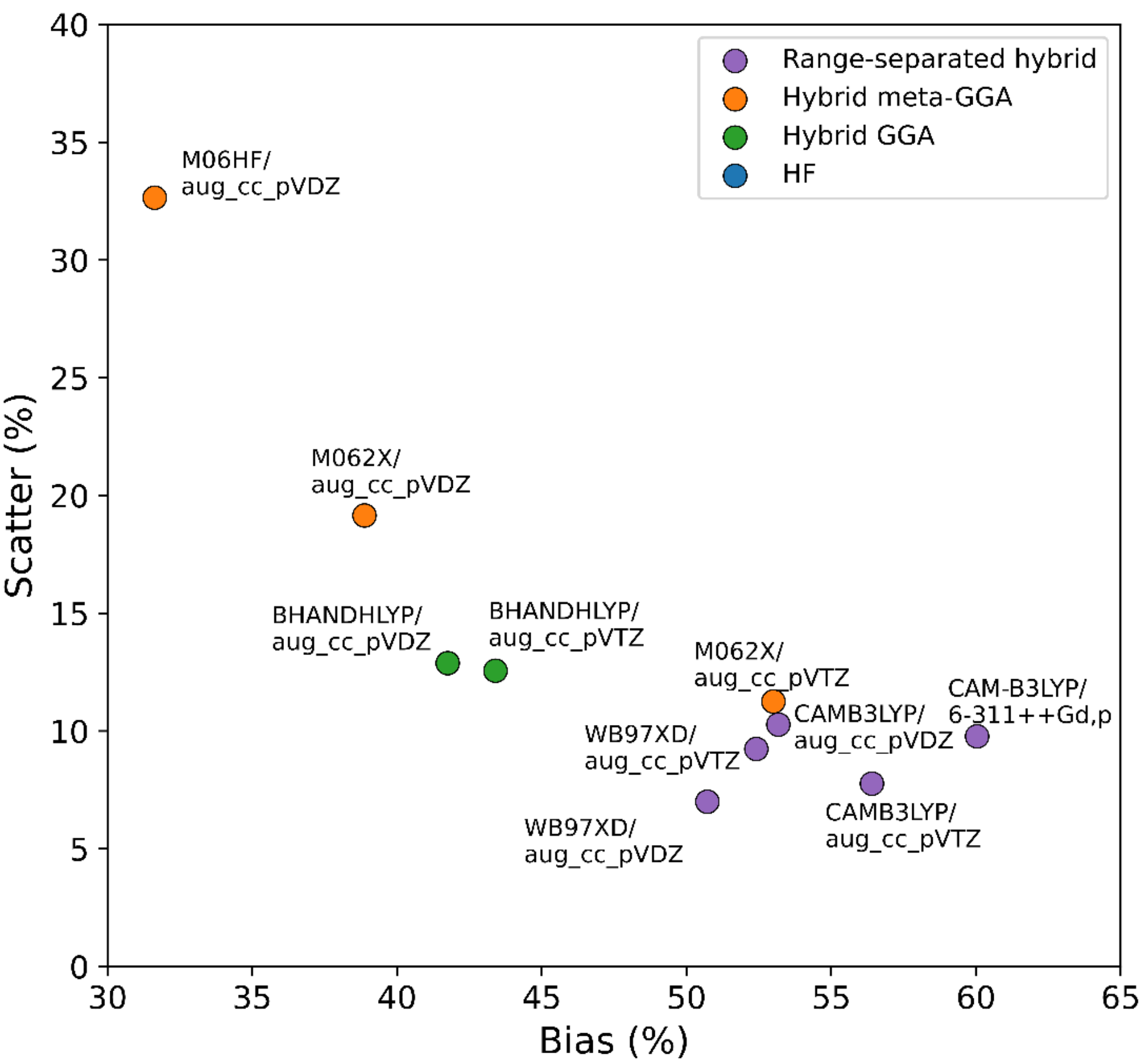

**Figure 4.** $d_{33}$ bias vs scatter across the experimental comparisons for the ground state geometry runs. Lower positions on the y-axis indicate a given method/basis set has a lower coefficient of variation of $\frac{d33_{calc}}{d33_{exp}}$. Its position on the x-axis shows the mean signed percentage error for the calculated ratios, $(\bar{r} - 1) * 100)$. Here, methods are evaluated based on their scatter as the bias is heavily influenced by the $\langle P_1 \rangle$ and $\langle P_3 \rangle$ order parameters.

We calculate $r = \frac{d33_{calc}}{d33_{exp}}$ for each calculation run and define two evaluation metrics from this, the bias and the scatter. Bias was defined as the mean signed percentage error across the three molecules, $((\bar{r} - 1) * 100)$ where $\bar{r}$ is the mean of the ratios. The bias quantifies any systematic over- or underprediction within the dataset. The scatter is taken to be the coefficient of variation of $r$, calculated as $\frac{\sigma_r}{\bar{r}}$ where $\sigma_r$ is the standard deviation of the ratios. This gives a scale-independent measure of the dispersion in the calculated-to-experimental ratios. This analysis should be taken at face value given the paucity of experimental data (n=3), but it does suggest a tendency for overprediction.

Figure 4 shows the bias-scatter plots for the ground-state geometry calculations. The bias shows that $d_{33}$ values are consistently overpredicted by all tested methods. The range-separated hybrids consistently produced low scattering values, indicating that they are most likely to successfully reproduce $r$, however they do all possess generally larger biases compared to BHandHLYP and M06-2X although wB97XD is smaller than M06-2X. The BHandHLYP method also performs well. Its scatter is slightly above the range-separated hybrid method but has a slightly lower bias. The scatter for hybrid meta-GGA methods shows a separation between M06-2X and M06-HF with a 13 percentage points decrease in the scattering for M06-2X. HF performs poorly in both scattering and bias when compared to the hybrid GGA and range-separated hybrid method as evidenced in Figure 1, indicating a deficiency in their ability to accurately reproduce the experimental value consistently. We can also see again that there is no real performance increase when using triple-zeta basis sets compared to the double-zeta sets. Based on the scatter and bias analysis in figure 4, we selected the wB97X-D/aug-cc-pVDZ method and basis set combination as it possessed the lowest scatter, indicating a consistent calculation of the $\frac{d33_{calc}}{d33_{exp}}$ ratios across our benchmarking data.

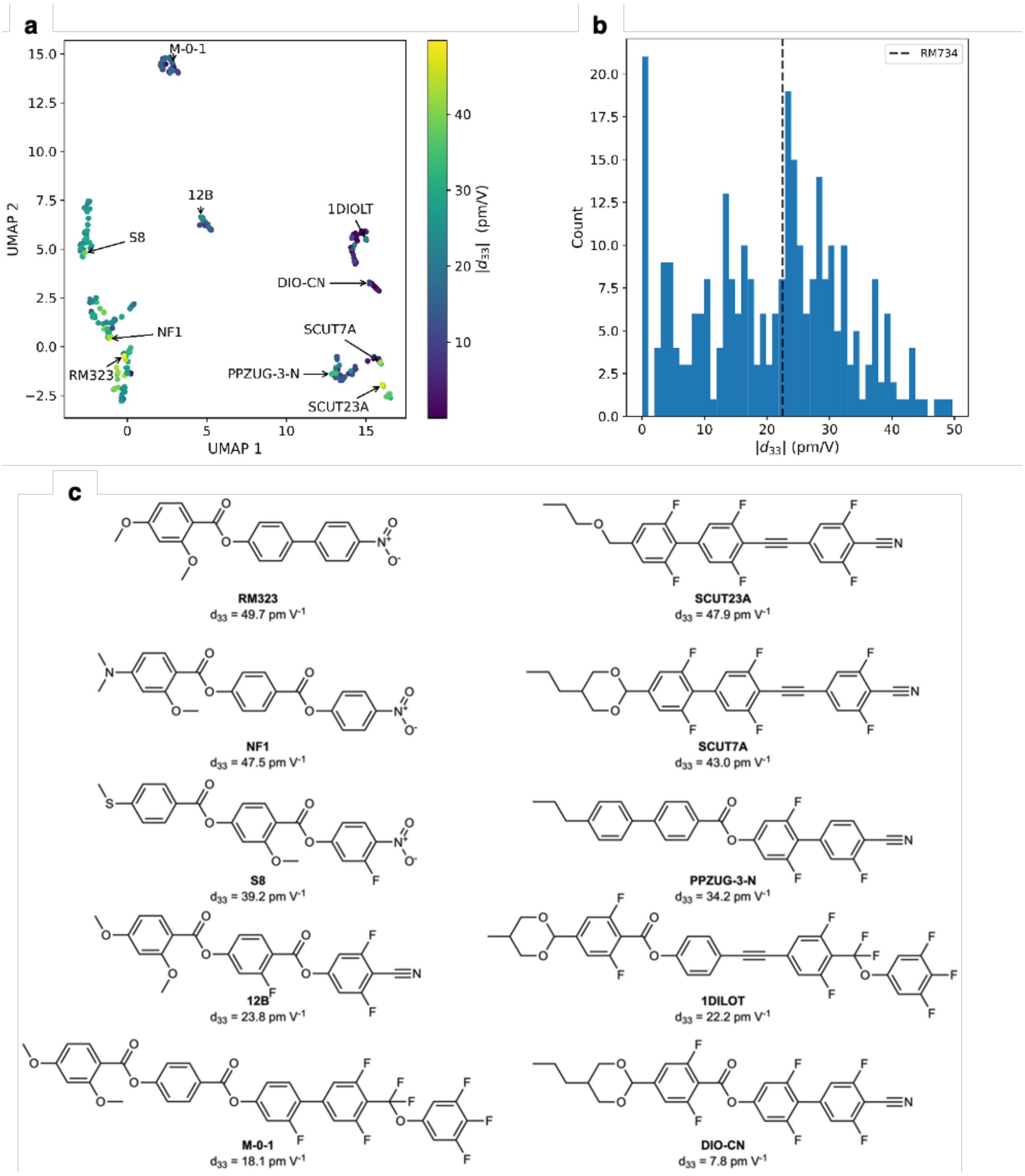

**Figure 5.** A) A UMAP projection of the ECFP4 fingerprints of the $N_F$ database, the data points are coloured by their $d_{33}$ magnitude. B) The distribution of calculated $d_{33}$ magnitudes across the $N_F$ database. The experimental value of RM734 is given for comparison purposes. C) Candidates for SHG screened from the $N_F$ database. The $d_{33}$ values are calculated at wB97X-D/aug-cc-pVDZ.

Having shown that single conformers yield acceptable results, and triple-zeta basis sets are hardly more performant than double-zeta (within the assumptions made), we computed hyperpolarisability tensors and NLO coefficients for a set of $N_F$ materials taken from the literature (see Methods) using a range of DFT functionals (wB97X-D, M06-HF, M06-2X, CAM-B3LYP) with the aug-cc-pVDZ basis set. The calculation of NLO coefficients from $\beta$ used the same parameters for density and odd-rank order parameters as used for RM734.

Figure 5a shows the UMAP projection of ECFP4 fingerprints for the $N_F$ dataset [53], colour coded by the value predicted by wB97X-D/aug-cc-pVDZ. Figure 5b shows the distribution of calculated $d_{33}$ coefficients across the $N_F$ dataset for wB97X-D/aug-cc-pVDZ with values ranging from 0 $pm/V$ to 49.7 $pm/V$. At this level RM734 is calculated to be 29.7 $pm/V$, versus an experimental value of 22.5 $pm/V$. Figure 5c presents notably performant examples from the dataset; the names can be used to locate their position in the UMAP space in Figure 5a. These examples were selected as they have the highest d coefficient within their UMAP cluster.

The macroscopic nonlinear optical coefficients depend not only on the molecular hyperpolarisability but also on material-specific parameters, including the odd-rank order parameters, molecular number density, and refractive indices. These are bulk properties and cannot meaningfully be determined from electronic structure calculations on isolated molecules, and are expected to show some variance between materials. Together, these likely contribute to the deviation between calculated and experimental d coefficients. As an objective of the present work is to compare the intrinsic nonlinear optical potential of many molecular candidates, the use of representative values of the order parameters and refractive indices (taken for RM734) normalises the contribution of molecular hyperpolarisability to the calculated nonlinear coefficients and provides a consistent basis for ranking molecular performance. In effect, the calculated coefficients should therefore be interpreted as relative screening metrics.

**Figure 6.** Molecular structures of reported $N_F$ materials with their calculated $d_{33}$ coefficients at the wB97XD/aug-cc-pVDZ level of DFT for the global low-energy geometry. The structures are grouped to highlight the impact of different donor, linking and lateral structural groups on the resulting $d_{33}$ coefficients.

The largest values of $\beta_{xxx}$ – and therefore $d_{33}$ – are consistently found in systems with a "push-pull" electron system. Although not designed as an NLO chromophore, RM734 features such a system with the 2,4-dimethoxybenzoate as a donor ("push") and the 4-nitrophenyl unit as an acceptor ("pull") and has a correspondingly large value of $d_{33}$. Unsurprisingly, using units such as N,N-dimethylamino (**NF1** [67]) or methylthio (**S8** [108]) gives larger $\beta_{xxx}$ as these are stronger donors, and correspondingly larger $d$ coefficients. Our calculations show that 2,5-disubstituted 1,3-dioxanes are deadweight in terms of NLO properties; they are poor donor groups and only give substantial calculated responses in combination with other functionalities, such as alkynes as in **SCUT5a** [109].

In a similar vein, trioxabicyclo[2.2.2]octanes are only effective when adjacent to the phenylacetylene motif (e.g. **3BOE** [110]); this has comparable NLO coefficients to the dioxane material **SCUT5a**, but these materials are likely to have issues around toxicity (EBOB is a potent GABA receptor agonist) [111]. Not all linking groups are created equal. Acetylenes outperform esters significantly, while esters, thioesters and direct phenyl-phenyl bonds are comparable. The difluoromethyleneoxy (Q-bridge; e.g. DUUQU-3-N) invariably leads to small values of $\beta$ due to its interruption of conjugation, an effect we see in well-known named $N_F$ materials from Merck UUQU-4-N ($d_{33} = 4.3\ pm/V$) [48] and AUUQU-2-N ($d_{33} = 3.7\ pm/V$) [112].

The utility of fluorine as a lateral group is well established in terms of its ability to enhance thermal stability of polar order [113]; however, this comes at a cost to the calculated NLO coefficients as shown in Figure 6. As nitro is already a very strong acceptor group, its replacement typically leads to smaller calculated NLO coefficients (for example, compare **12B** with RM734). While not in the $N_F$ dataset here, incorporating typical NLO-like acceptors such as dicyanovinylene as polar terminal groups may be one route to retaining $N_F$ phases and with even larger NLO coefficients. Similarly, extending the $\pi$-conjugated system in the molecule is associated with increased NLO coefficients, either by adding additional aromatic rings or using conjugated linkers (e.g. acetylenes).

Together, the resulting design rule for $N_F$ materials with large second-order NLO coefficients is rather simple – a strong donor-acceptor asymmetry connected by an extended, planar, conjugated pathway along the molecular long axis. Motifs that promote polar organisation, but interrupt conjugation (e.g. difluoromethyleneoxy) are inevitably detrimental in this regard.

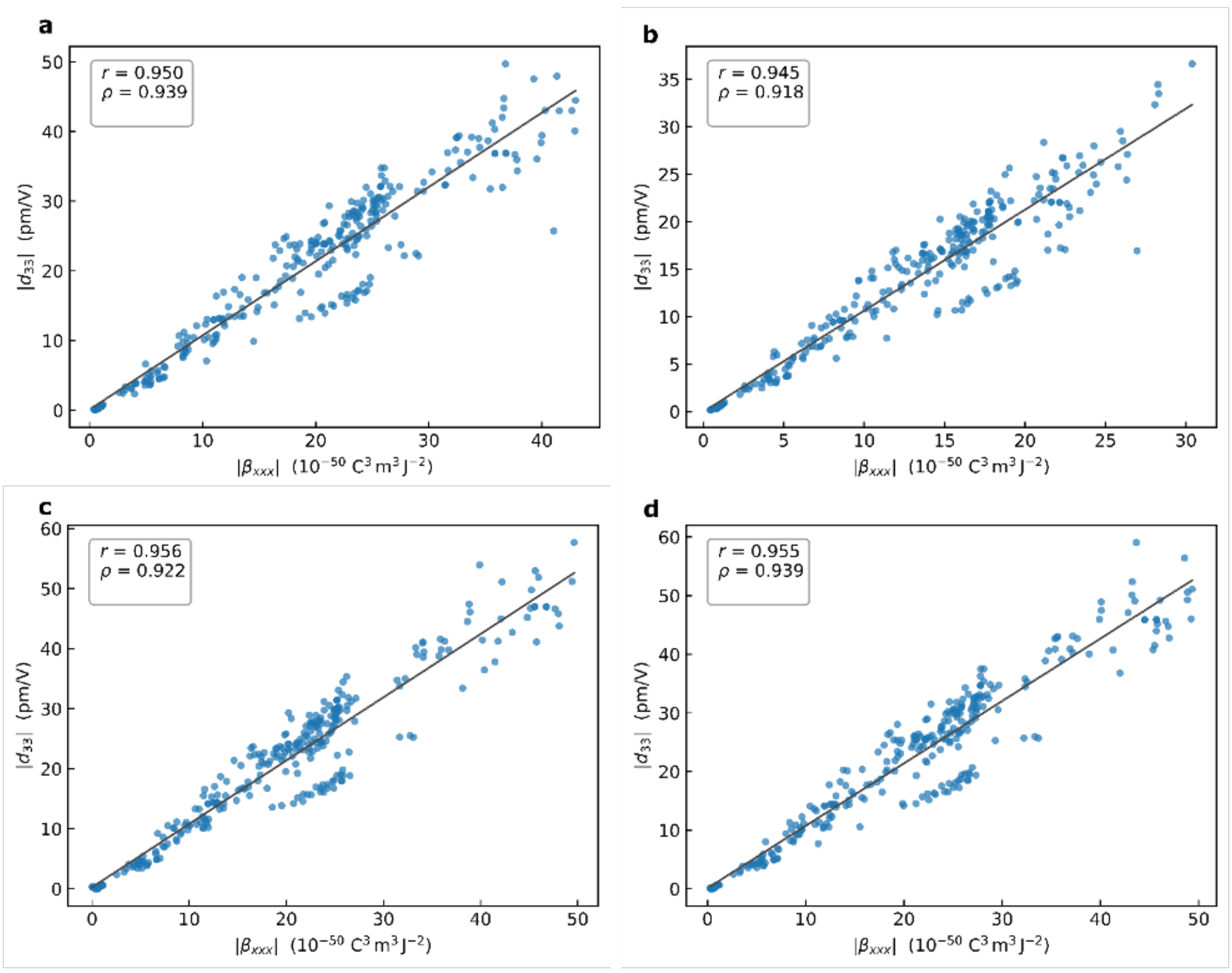

**Figure 7.** Scatter plots of $|\beta_{xxx}|$ against $|d_{33}|$ for the full dataset using the aug-cc-pVDZ basis set and the following DFT functionals: (a) wB97X-D; (b) M06-HF; (c) M06-2X; (d) CAM-B3LYP

Figure 7 shows plots of $|\beta_{xxx}|$ against $|d_{33}|$ for the selected DFT functionals using the aug-cc-pVDZ basis set. As the calculation uses the assumed parameters (field factor calculated from refractive indices, density, $\langle P_1 \rangle$ and $\langle P_3 \rangle$ order parameter values) the calculation essentially becomes a linear rescaling of the $\beta_{xxx}$ with an influence from the $\beta_{yyx}, \beta_{zzx}, \beta_{xxy}, \beta_{xzz}$ tensor components along with the molecular mass; indeed, there is a linear relation between $\beta_{xxx}$ and $d_{33}$. Consequently, we suggest that when identifying candidates for SHG, further consideration must be given to the material's bulk behaviour. Experimental values of bulk density and odd-rank order parameters for polar liquid crystals are almost entirely unknown. If the odd-rank order parameters and material density vary only within narrow ranges, data-driven models, such as those recently reported for dielectric anisotropy [114], may outperform physics-based approximations, given sufficient input data. However, they may vary significantly between materials. Especially in the case of the odd-rank order parameters, these would be expected display a dependence on temperature (and upon the phase sequence in question).

## Conclusion

We have established a workflow for calculating the second-order nonlinear optical coefficients of polar liquid crystals *via* electronic structure calculations. There is no standout method, but wB97X-D, M06-HF, M06-2X and CAM-B3LYP all perform adequately when using a Dunning-type correlation-consistent double-zeta basis set such as aug-cc-pVDZ with wB97X-D values presented for the screening process. We find that explicit low-energy conformer ensembles are unnecessary for screening purposes with the chosen methods. Instead, reasonable estimates for the NLO coefficients can be obtained from the global-minimum-energy geometry using relatively inexpensive DFT functionals and a double-zeta basis set. We use these methods to screen a dataset of known $N_F$ materials, allowing us to identify basic design principles for generating large NLO coefficients in ferroelectric nematic materials.

## Code and Data Availability

Code and data relevant to this article are available *via* GitHub at https://github.com/123cpb/hp2nlo

## Acknowledgements

R.J.M. thanks UKRI for funding via a Future Leaders Fellowship, grant number MR/W006391/1, and the University of Leeds for funding via a University Academic Fellowship. N.S. acknowledges the financial support of the Slovenian Research Agency (Grant Nos. P1-0192, J1-50004, and BI-VB/25-27-011)

Supplementary Information for:

# Molecular Hyperpolarisability as a Screening Descriptor for Second-Order Nonlinear Optics in Ferroelectric Nematic Liquid Crystals

Charles Parton-Barr[1], Nerea Sebastian[2], Richard J. Mandle [1,3] *

[1]School of Physics and Astronomy, University of Leeds, Leeds, UK, LS2 9JT
[2] Complex Matter Group, Jozef Stefan Institut, Ljubljana, Slovenia
[3] School of Chemistry, University of Leeds, Leeds, UK, LS2 9JT

**Contents**

## 1. Minimum energy vs conformer comparison

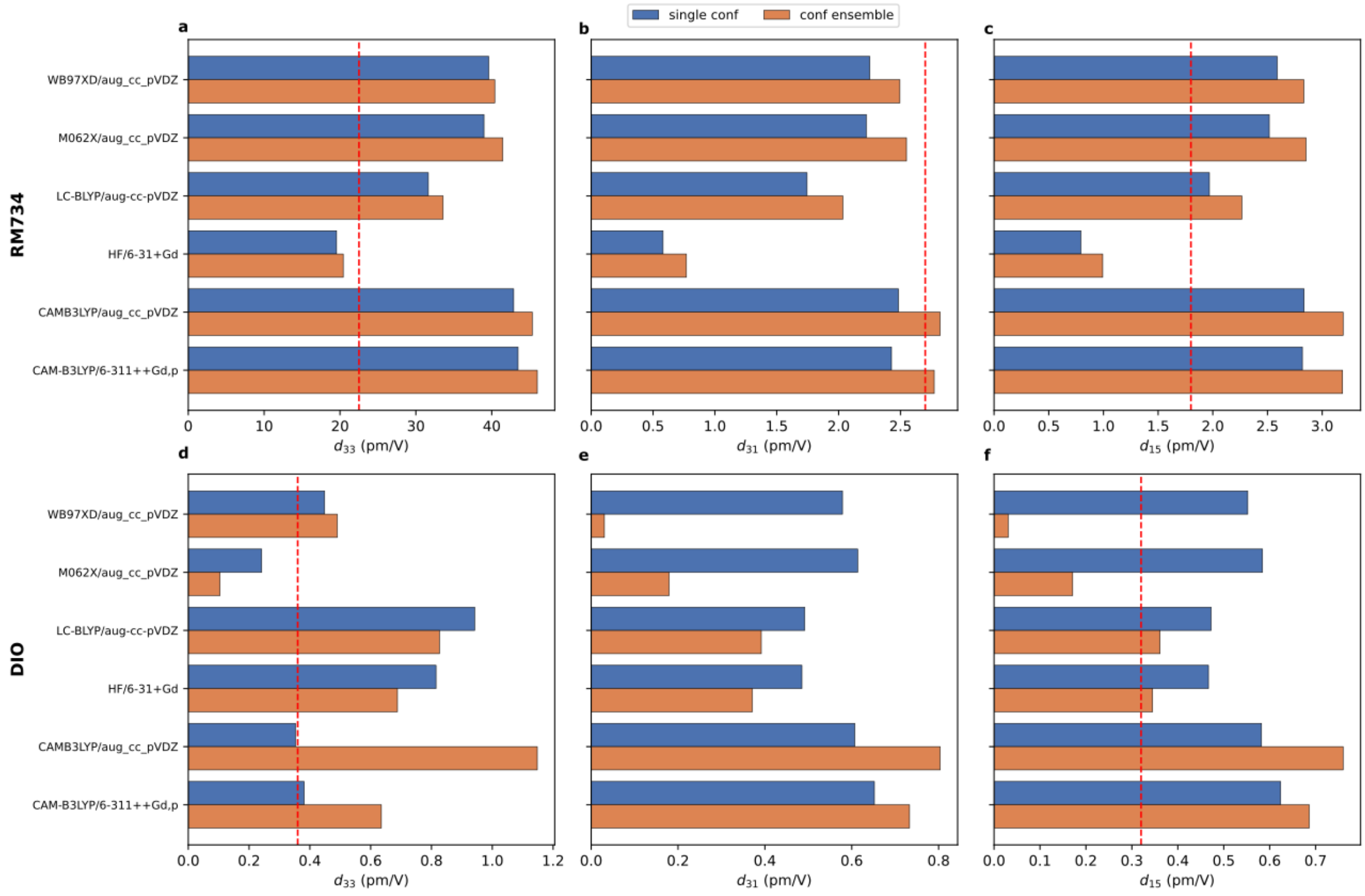

**Figure SI – 1.** A comparison between the between the single conformer and conformational ensemble calculated d coefficients for RM734 (a-c) and DIO (d-f)

2. **Conformer ensemble d33 ratio**

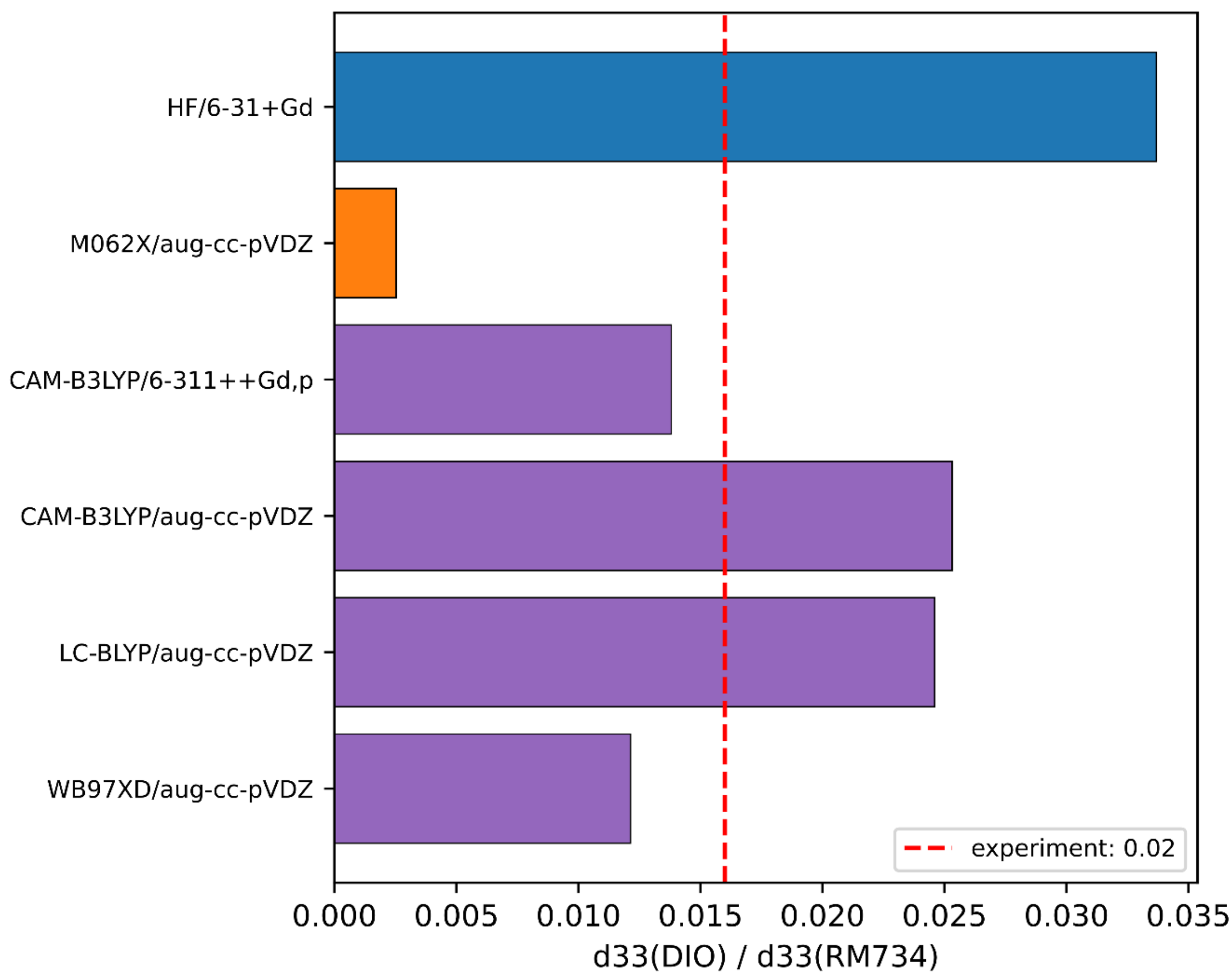

**Figure SI-2:** Ratios of calculated d33 in the conformational ensemble condition between RM734 and DIO.

## 3. Calculated $d_{il}$ coefficients for benchmarks

```
================================================================================
Benchmark Results: RM734
Extraction wavelength: 800.0 nm
================================================================================
Method/Basis                              d33         d31          d15
--------------------------------------------------------------------------------
Experimental                            22.50         2.7          1.8
--------------------------------------------------------------------------------
B3LYP/aug_cc_pVDZ                       82.46        5.29         5.92
B3LYP/aug_cc_pVTZ                       80.14        5.12         5.75
BHANDHLYP/aug_cc_pVDZ                   36.50        1.94         2.27
BHANDHLYP/aug_cc_pVTZ                   35.92        1.91         2.23
CAM-B3LYP/6-311++Gd,p                   43.46        2.43         2.82
CAMB3LYP/aug_cc_pVDZ                    42.89        2.48         2.83
CAMB3LYP/aug_cc_pVTZ                    42.26        2.45         2.79
HF/6-31+Gd                              19.54        0.58         0.79
HF/aug_cc_pVDZ                          19.77        0.64         0.84
HF/aug_cc_pVTZ                          19.51        0.63         0.83
LC-BLYP/aug-cc-pVDZ                     31.64        1.74         1.97
M062X/aug_cc_pVDZ                       38.97        2.22         2.52
M062X/aug_cc_pVTZ                       38.42        2.21         2.50
M06/AUG_CC_PVDZ                         64.56        3.82         4.44
M06/AUG_CC_PVTZ                         65.29        4.00         4.57
M06HF/aug_cc_pVDZ                       26.75        1.41         1.57
M06HF/aug_cc_pVTZ                       25.31        1.27         1.44
TPSSH/aug_cc_pVDZ                      322.66       23.46        24.36
TPSSH/aug_cc_pVTZ                      196.02       13.91        14.70
WB97XD/aug_cc_pVDZ                      39.60        2.25         2.59
WB97XD/aug_cc_pVTZ                      39.16        2.25         2.58
================================================================================
Best d33 match: HF/aug_cc_pVDZ (error: -12.1%)
```

**Figure SI-3:** Calculated d coefficients for single conformers of RM734

```
================================================================================
Benchmark Results: RM734
Extraction wavelength: 800.0 nm
================================================================================
Method/Basis                         d33          d31          d15
--------------------------------------------------------------------------------
Experimental                       22.50          2.7          1.8
--------------------------------------------------------------------------------
CAM-B3LYP/6-311++Gd,p             -46.00        -2.77        -3.19
CAM-B3LYP/aug-cc-pVDZ             -45.34        -2.82        -3.19
HF/6-31+Gd                        -20.42        -0.77        -0.99
LC-BLYP/aug-cc-pVDZ               -33.59        -2.03        -2.27
M062X/aug-cc-pVDZ                 -41.45        -2.55        -2.85
WB97XD/aug-cc-pVDZ                -40.40        -2.50        -2.83
================================================================================
Best d33 match: HF/6-31+Gd (error: -9.2%)
```

**Figure SI-4:** Calculated d coefficients for a conformational ensemble of RM734

```
================================================================================
Benchmark Results: DIO
Extraction wavelength: 800.0 nm
================================================================================
Method/Basis                         d33          d31          d15
--------------------------------------------------------------------------------
Experimental                        0.36          N/A         0.32
--------------------------------------------------------------------------------
B3LYP/aug_cc_pVDZ                  -2.98        -0.94        -0.92
B3LYP/aug_cc_pVTZ                  -2.94        -0.92        -0.90
BHANDHLYP/aug_cc_pVDZ               0.35        -0.57        -0.55
BHANDHLYP/aug_cc_pVTZ               0.39        -0.55        -0.53
CAM-B3LYP/6-311++Gd,p               0.38        -0.65        -0.62
CAMB3LYP/aug_cc_pVDZ                0.35        -0.61        -0.58
CAMB3LYP/aug_cc_pVTZ                0.41        -0.59        -0.56
HF/6-31+Gd                          0.81        -0.49        -0.47
HF/aug_cc_pVDZ                      0.79        -0.47        -0.45
HF/aug_cc_pVTZ                      0.83        -0.45        -0.44
LC-BLYP/aug-cc-pVDZ                 0.94        -0.49        -0.47
M062X/aug_cc_pVDZ                   0.24        -0.61        -0.58
M062X/aug_cc_pVTZ                   0.45        -0.54        -0.52
M06/AUG_CC_PVDZ                    -1.71        -0.84        -0.82
M06/AUG_CC_PVTZ                    -1.64        -0.80        -0.78
M06HF/aug_cc_pVDZ                   0.54        -0.58        -0.54
M06HF/aug_cc_pVTZ                   0.71        -0.50        -0.47
TPSSH/aug_cc_pVDZ                  -6.08        -1.18        -1.16
TPSSH/aug_cc_pVTZ                  -6.07        -1.16        -1.15
WB97XD/aug_cc_pVDZ                  0.45        -0.58        -0.55
WB97XD/aug_cc_pVTZ                  0.48        -0.57        -0.54
================================================================================
Best d33 match: CAMB3LYP/aug_cc_pVDZ (error: -2.0%)
```

**Figure SI-5**: Calculated d coefficients for single conformers of DIO

```
================================================================================
Benchmark Results: DIO
Extraction wavelength: 800.0 nm
================================================================================
Method/Basis                          d33          d31          d15
--------------------------------------------------------------------------------
Experimental                         0.36          N/A         0.32
--------------------------------------------------------------------------------
CAM-B3LYP/6-311++Gd,p                0.64         0.73         0.69
CAM-B3LYP/aug-cc-pVDZ                1.15         0.80         0.76
HF/6-31+Gd                          -0.69         0.37         0.35
LC-BLYP/aug-cc-pVDZ                 -0.83         0.39         0.36
M062X/aug-cc-pVDZ                   -0.10         0.18         0.17
WB97XD/aug-cc-pVDZ                  -0.49        -0.03        -0.03
================================================================================
Best d33 match: WB97XD/aug-cc-pVDZ (error: +36.1%)
```

**Figure SI-6:** Calculated d coefficients for a conformational ensemble of DIO

```
================================================================================
Benchmark Results: C1
Extraction wavelength: 800.0 nm
================================================================================
Method/Basis                          d33          d31          d15
--------------------------------------------------------------------------------
Experimental                         3.90          N/A         0.35
--------------------------------------------------------------------------------
B3LYP/aug_cc_pVDZ                   12.74        -0.70        -0.62
B3LYP/aug_cc_pVTZ                   12.69        -0.69        -0.61
BHANDHLYP/aug_cc_pVDZ                8.04        -0.63        -0.57
BHANDHLYP/aug_cc_pVTZ                8.01        -0.62        -0.56
CAM-B3LYP/6-311++Gd,p                8.81        -0.71        -0.65
CAMB3LYP/aug_cc_pVDZ                 8.42        -0.66        -0.60
CAMB3LYP/aug_cc_pVTZ                 8.43        -0.65        -0.59
HF/6-31+Gd                           5.06        -0.64        -0.59
HF/aug_cc_pVDZ                       4.87        -0.61        -0.57
LC-B3LYP/AUG_CC_PVDZ                 6.35        -0.61        -0.56
M062X/aug_cc_pVDZ                    8.24        -0.65        -0.59
M062X/aug_cc_pVTZ                    8.35        -0.60        -0.54
M06/AUG_CC_PVDZ                     11.82        -0.71        -0.62
M06HF/aug_cc_pVDZ                    6.08        -0.64        -0.59
TPSSH/aug_cc_pVDZ                   14.30        -0.68        -0.60
TPSSH/aug_cc_pVTZ                   14.08        -0.68        -0.59
WB97XD/aug_cc_pVDZ                   7.67        -0.67        -0.61
WB97XD/aug_cc_pVTZ                   7.68        -0.65        -0.60
================================================================================
Best d33 match: HF/aug_cc_pVDZ (error: +24.9%)
```

**Figure SI-7:** Calculated d coefficients for single conformers of C1

## 4. Calculated $d_{il}$ coefficients per molecule at the wB97xd/aug-cc-pvDZ level

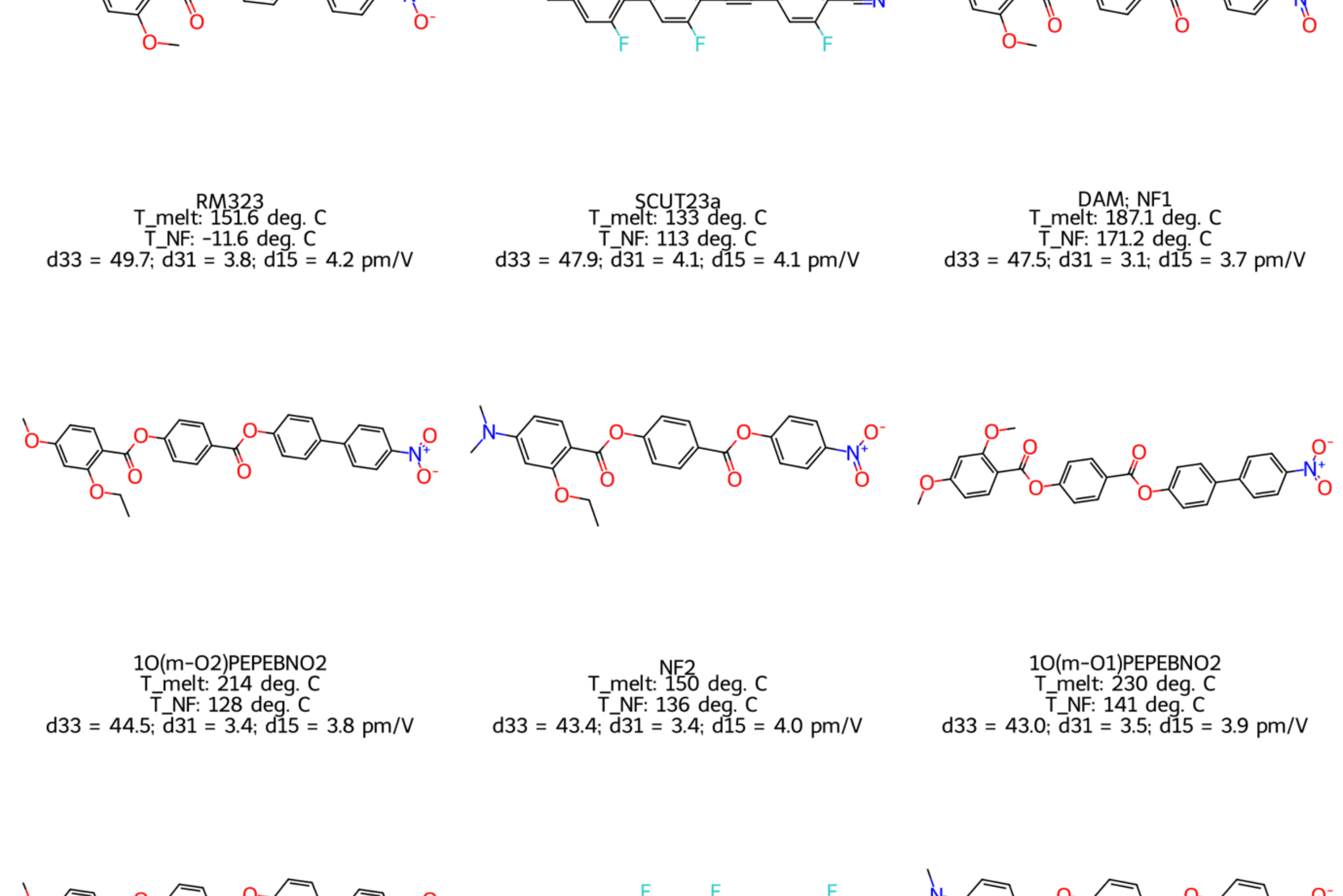

RM323
T_melt: 151.6 deg. C
T_NF: -11.6 deg. C
d33 = 49.7; d31 = 3.8; d15 = 4.2 pm/V

SCUT23a
T_melt: 133 deg. C
T_NF: 113 deg. C
d33 = 47.9; d31 = 4.1; d15 = 4.1 pm/V

DAM; NF1
T_melt: 187.1 deg. C
T_NF: 171.2 deg. C
d33 = 47.5; d31 = 3.1; d15 = 3.7 pm/V

1O(m-O2)PEPEBNO2
T_melt: 214 deg. C
T_NF: 128 deg. C
d33 = 44.5; d31 = 3.4; d15 = 3.8 pm/V

NF2
T_melt: 150 deg. C
T_NF: 136 deg. C
d33 = 43.4; d31 = 3.4; d15 = 4.0 pm/V

1O(m-O1)PEPEBNO2
T_melt: 230 deg. C
T_NF: 141 deg. C
d33 = 43.0; d31 = 3.5; d15 = 3.9 pm/V

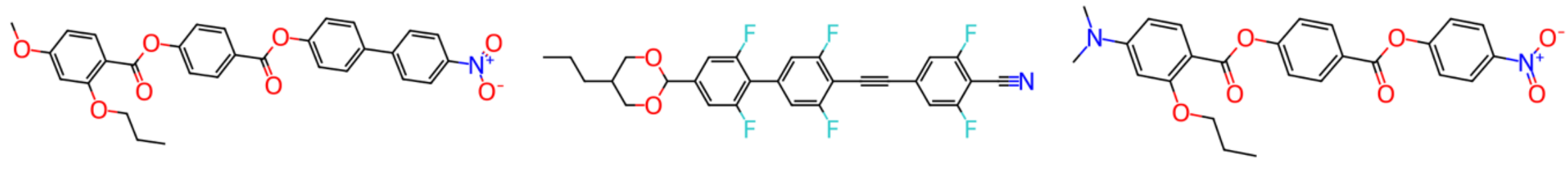

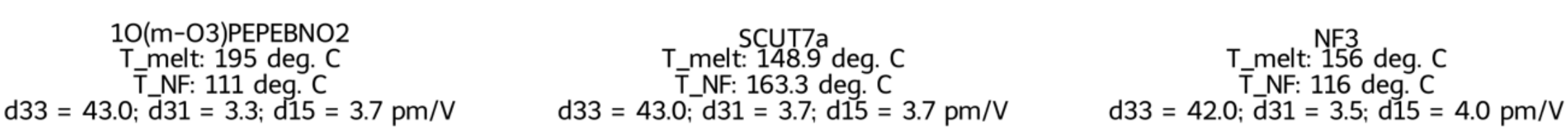

1O(m-O3)PEPEBNO2
T_melt: 195 deg. C
T_NF: 111 deg. C
d33 = 43.0; d31 = 3.3; d15 = 3.7 pm/V

SCUT7a
T_melt: 148.9 deg. C
T_NF: 163.3 deg. C
d33 = 43.0; d31 = 3.7; d15 = 3.7 pm/V

NF3
T_melt: 156 deg. C
T_NF: 116 deg. C
d33 = 42.0; d31 = 3.5; d15 = 4.0 pm/V

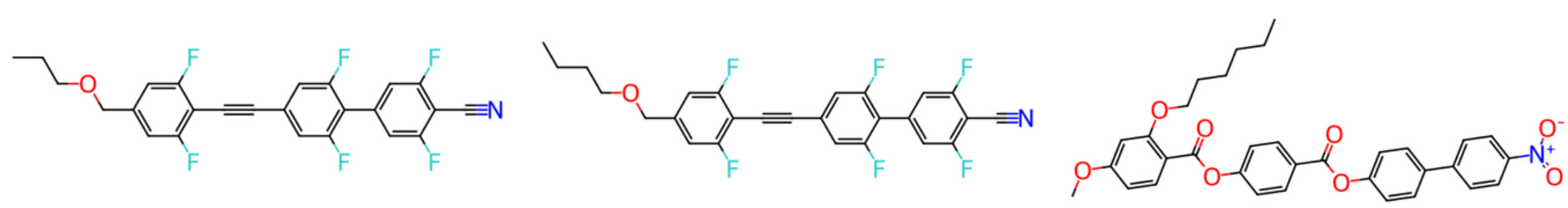

SCUT22b
T_melt: 122 deg. C
T_NF: 106.9 deg. C
d33 = 41.2; d31 = 3.4; d15 = 3.4 pm/V

SCUT22c
T_melt: 98.2 deg. C
T_NF: 80.5 deg. C
d33 = 40.3; d31 = 3.4; d15 = 3.5 pm/V

1O(m-O6)PEPEBNO2
T_melt: 150 deg. C
T_NF: 70 deg. C
d33 = 40.1; d31 = 3.1; d15 = 3.5 pm/V

**Figure 7:** Molecular structure, transition temperatures ($°C$) and calculated $d_{il}$ coefficients ($pm/V$) for materials exhibiting an $N_F$ phase. Transition temperatures and molecular structures were taken from refs [1-47]

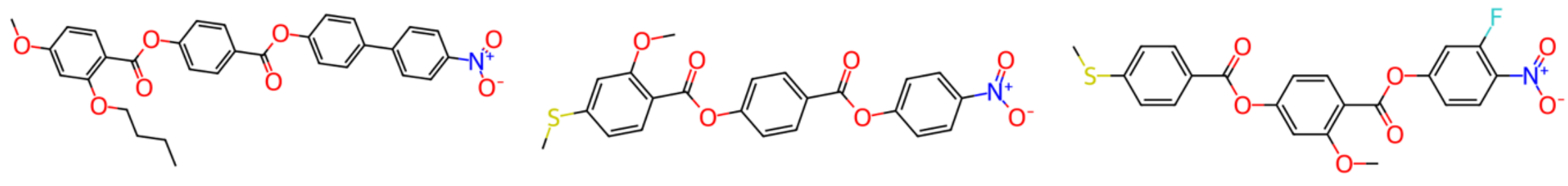

1O(m-O4)PEPEBNO2
T_melt: 153 deg. C
T_NF: 95 deg. C
d33 = 39.4; d31 = 3.2; d15 = 3.6 pm/V

2
T_melt: 148.7 deg. C
T_NF: 105.8 deg. C
d33 = 39.4; d31 = 2.8; d15 = 3.2 pm/V

S8
T_melt: 164 deg. C
T_NF: 124 deg. C
d33 = 39.2; d31 = 2.8; d15 = 3.2 pm/V

7; GS36
T_melt: 173.3 deg. C
T_NF: 106 deg. C
d33 = 39.1; d31 = 2.8; d15 = 3.2 pm/V

NF4
T_melt: 144 deg. C
T_NF: 96 deg. C
d33 = 39.0; d31 = 4.0; d15 = 4.5 pm/V

SCUT22d
T_melt: 78 deg. C
T_NF: 60 deg. C
d33 = 38.7; d31 = 3.5; d15 = 3.5 pm/V

1O(m-O5)PEPEBNO2
T_melt: 156 deg. C
T_NF: 82 deg. C
d33 = 38.4; d31 = 3.2; d15 = 3.6 pm/V

NF5
T_melt: 120 deg. C
T_NF: 82 deg. C
d33 = 37.7; d31 = 3.4; d15 = 4.0 pm/V

S2
T_melt: 147 deg. C
T_NF: 122 deg. C
d33 = 37.4; d31 = 2.5; d15 = 2.9 pm/V

S4
T_melt: 150 deg. C
T_NF: 129 deg. C
d33 = 37.0; d31 = 2.3; d15 = 2.7 pm/V

S3
T_melt: 162 deg. C
T_NF: 126 deg. C
d33 = 37.0; d31 = 3.0; d15 = 3.4 pm/V

BIOTN2; 2BOE
T_melt: 168.4 deg. C
T_NF: 289 deg. C
d33 = 36.9; d31 = 2.9; d15 = 2.9 pm/V

**Figure SI-7:** Continued.

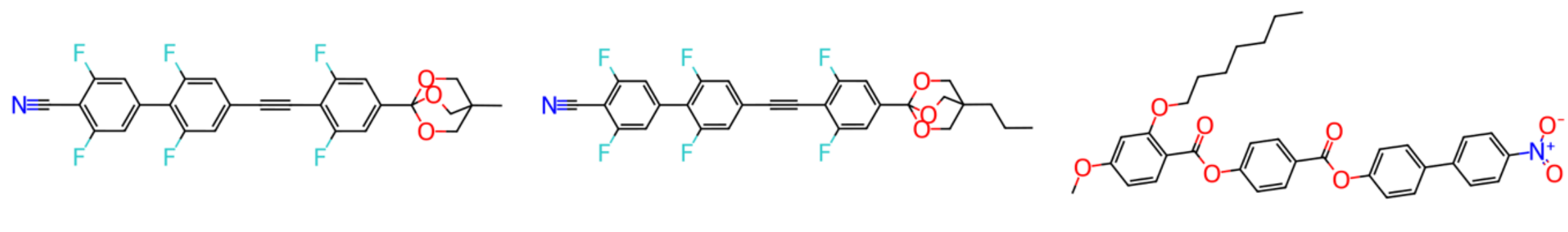

BIOTN1; 1BOE
T_melt: 192.5 deg. C
T_NF: 296.1 deg. C
d33 = 36.9; d31 = 2.8; d15 = 2.9 pm/V

3BOE
T_melt: 190 deg. C
T_NF: 280 deg. C
d33 = 36.7; d31 = 2.9; d15 = 2.9 pm/V

1O(m-O7)PEPEBNO2
T_melt: 164 deg. C
T_NF: 69 deg. C
d33 = 36.0; d31 = 2.8; d15 = 3.1 pm/V

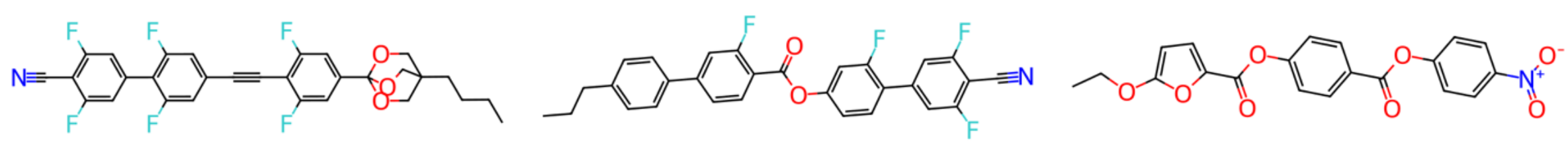

4BOE
T_melt: 155 deg. C
T_NF: 245 deg. C
d33 = 36.0; d31 = 2.8; d15 = 2.9 pm/V

CJG296
T_melt: 137.3 deg. C
T_NF: N/A
d33 = 35.5; d31 = 2.6; d15 = 2.9 pm/V

CJG-1-11
T_melt: 191.4 deg. C
T_NF: 205 deg. C
d33 = 34.8; d31 = 3.0; d15 = 3.4 pm/V

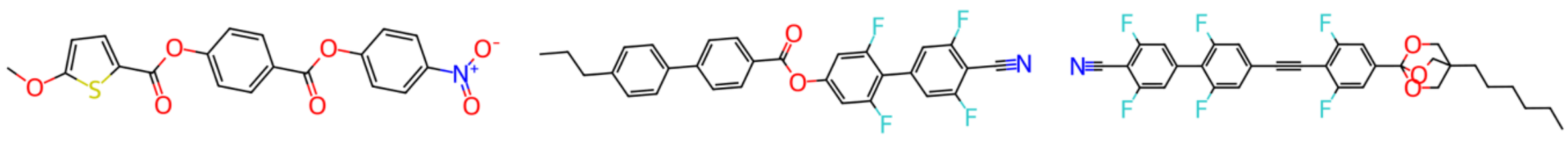

15
T_melt: 188.0 deg. C
T_NF: 161.1 deg. C
d33 = 34.8; d31 = 2.6; d15 = 3.1 pm/V

CJG486
T_melt: 162.2 deg. C
T_NF: 122.3 deg. C
d33 = 34.4; d31 = 2.6; d15 = 2.9 pm/V

6BOE
T_melt: 175 deg. C
T_NF: 180 deg. C
d33 = 34.4; d31 = 2.8; d15 = 2.9 pm/V

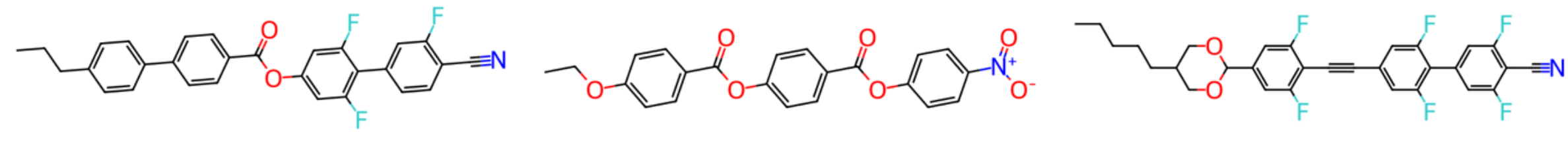

CJG350
T_melt: 134.6 deg. C
T_NF: N/A
d33 = 34.2; d31 = 2.7; d15 = 3.0 pm/V

RM63
T_melt: 155.2 deg. C
T_NF: -17.4 deg. C
d33 = 33.7; d31 = 2.5; d15 = 2.9 pm/V

SCUT5b
T_melt: 108.4 deg. C
T_NF: 96.9 deg. C
d33 = 33.4; d31 = 3.0; d15 = 3.1 pm/V

**Figure SI-7:** Continued.

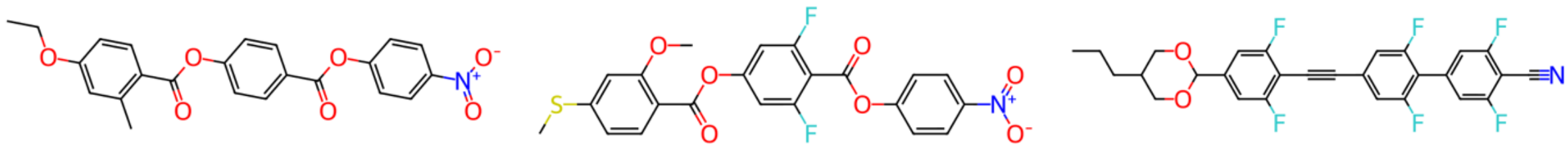

RM569
T_melt: 138.8 deg. C
T_NF: 32.7 deg. C
d33 = 32.9; d31 = 2.3; d15 = 2.7 pm/V

S5
T_melt: 174 deg. C
T_NF: 121 deg. C
d33 = 32.7; d31 = 1.9; d15 = 2.3 pm/V

AK3; SCUT5a
T_melt: 62.3 deg. C
T_NF: 185 deg. C
d33 = 32.3; d31 = 2.5; d15 = 2.5 pm/V

RM230; 11/01/2002
T_melt: 139 deg. C
T_NF: 85.6 deg. C
d33 = 32.1; d31 = 2.2; d15 = 2.5 pm/V

Aya_2A; 7-0-1; NT3.1
T_melt: 192 deg. C
T_NF: 125.6 deg. C
d33 = 32.1; d31 = 2.1; d15 = 2.5 pm/V

RM692
T_melt: 134.7 deg. C
T_NF: 8.7 deg. C
d33 = 32.1; d31 = 2.1; d15 = 2.5 pm/V

1ECMe
T_melt: 152 deg. C
T_NF: 71 deg. C
d33 = 32.0; d31 = 2.5; d15 = 2.9 pm/V

1O(m-O9)PEPEBNO2
T_melt: 188 deg. C
T_NF: 65 deg. C
d33 = 32.0; d31 = 2.8; d15 = 3.1 pm/V

5BOE-NO2
T_melt: 190.1 deg. C
T_NF: 212.8 deg. C
d33 = 31.8; d31 = 2.5; d15 = 2.5 pm/V

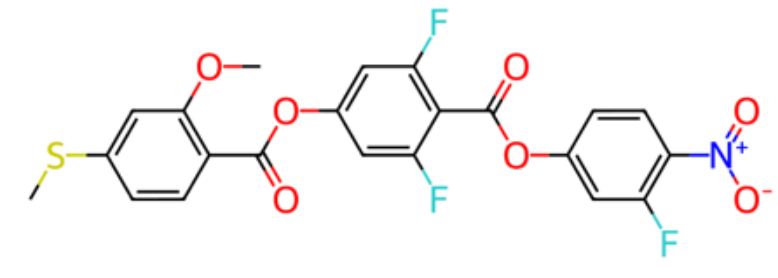

S6
T_melt: 171 deg. C
T_NF: 122 deg. C
d33 = 31.4; d31 = 1.8; d15 = 2.2 pm/V

CJG300
T_melt: 134.3 deg. C
T_NF: N/A
d33 = 31.4; d31 = 2.4; d15 = 2.6 pm/V

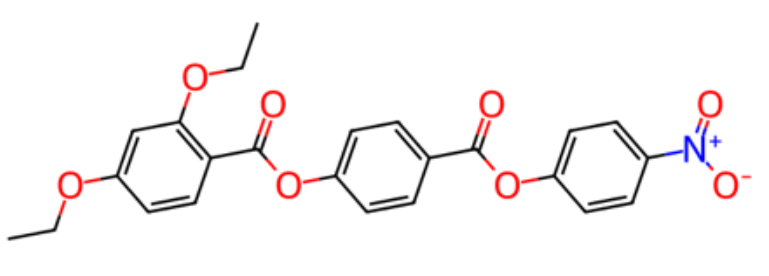

RM231; 11/02/2002
T_melt: 143.3 deg. C
T_NF: 91 deg. C
d33 = 31.4; d31 = 2.1; d15 = 2.4 pm/V

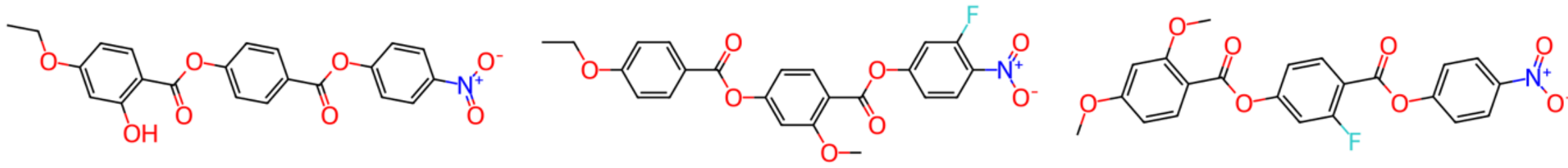

RM38
T_melt: 164.7 deg. C
T_NF: 1.8 deg. C
d33 = 31.1; d31 = 2.1; d15 = 2.5 pm/V

2OEC3F
T_melt: 168 deg. C
T_NF: 103 deg. C
d33 = 30.7; d31 = 2.1; d15 = 2.4 pm/V

RT11101; Imrie_2; O3; 28a
T_melt: 160.7 deg. C
T_NF: 140.5 deg. C
d33 = 30.6; d31 = 1.9; d15 = 2.3 pm/V

RT11165
T_melt: N/A
T_NF: 63 deg. C
d33 = 30.4; d31 = 2.0; d15 = 2.3 pm/V

RM1994_4; N1; 12/01/2001
T_melt: 165.2 deg. C
T_NF: 139.56 deg. C
d33 = 30.3; d31 = 1.9; d15 = 2.2 pm/V

M3; C3RM
T_melt: 147 deg. C
T_NF: 85 deg. C
d33 = 30.1; d31 = 1.9; d15 = 2.2 pm/V

1ECMeF
T_melt: 163 deg. C
T_NF: 117 deg. C
d33 = 30.1; d31 = 2.1; d15 = 2.4 pm/V

8-0-1; NT3F.1
T_melt: 180 deg. C
T_NF: 142 deg. C
d33 = 30.1; d31 = 2.0; d15 = 2.3 pm/V

EC1F
T_melt: 162 deg. C
T_NF: 144 deg. C
d33 = 29.9; d31 = 2.3; d15 = 2.6 pm/V

RM734; 11/01/2001
T_melt: 139.8 deg. C
T_NF: 132.7 deg. C
d33 = 29.7; d31 = 2.2; d15 = 2.6 pm/V

RM232
T_melt: 143.2 deg. C
T_NF: 77.6 deg. C
d33 = 29.6; d31 = 1.7; d15 = 2.1 pm/V

NT3F.2
T_melt: 179 deg. C
T_NF: 118 deg. C
d33 = 29.4; d31 = 1.9; d15 = 2.2 pm/V

**Figure SI-7:** Continued.

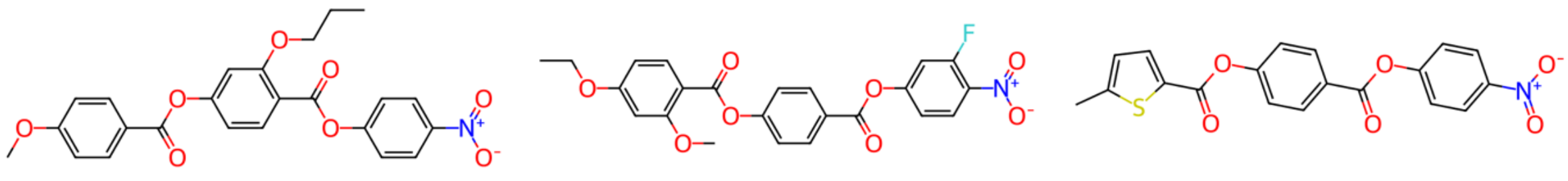

Aya_2C; NT3.3
T_melt: 86.8 deg. C
T_NF: 105.1 deg. C
d33 = 29.4; d31 = 2.0; d15 = 2.3 pm/V

RM554; 12/01/2002
T_melt: 141.8 deg. C
T_NF: 117.1 deg. C
d33 = 29.4; d31 = 2.1; d15 = 2.4 pm/V

14
T_melt: 183.4 deg. C
T_NF: 162.2 deg. C
d33 = 29.3; d31 = 2.0; d15 = 2.5 pm/V

Aya_2B; NT3.2
T_melt: 73.2 deg. C
T_NF: 96.8 deg. C
d33 = 29.2; d31 = 2.1; d15 = 2.4 pm/V

27a; EC3F
T_melt: 150 deg. C
T_NF: 158 deg. C
d33 = 29.0; d31 = 2.0; d15 = 2.3 pm/V

NT3.5
T_melt: 102 deg. C
T_NF: 64 deg. C
d33 = 28.7; d31 = 2.1; d15 = 2.4 pm/V

I-2; 11/02/2001
T_melt: 159 deg. C
T_NF: 106 deg. C
d33 = 28.6; d31 = 2.5; d15 = 2.8 pm/V

NT3F.3
T_melt: 135 deg. C
T_NF: 92 deg. C
d33 = 28.4; d31 = 1.8; d15 = 2.1 pm/V

N3
T_melt: 129 deg. C
T_NF: 88 deg. C
d33 = 28.4; d31 = 1.8; d15 = 2.1 pm/V

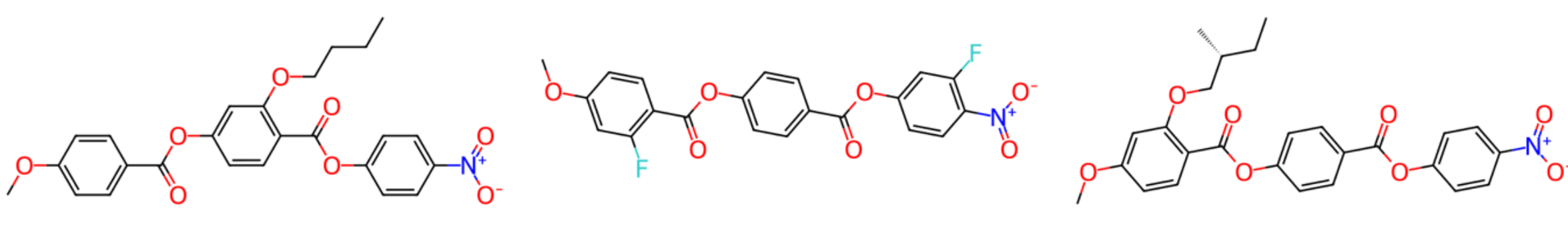

NT3.4
T_melt: 120 deg. C
T_NF: 76 deg. C
d33 = 28.3; d31 = 2.0; d15 = 2.3 pm/V

EC2F
T_melt: 155 deg. C
T_NF: 165 deg. C
d33 = 28.3; d31 = 2.0; d15 = 2.3 pm/V

I-4*
T_melt: 140 deg. C
T_NF: 54.8 deg. C
d33 = 27.9; d31 = 1.8; d15 = 2.1 pm/V

**Figure SI-7:** Continued.

Aya_1H
T_melt: N/A
T_NF: 76.2 deg. C
d33 = 27.9; d31 = 1.8; d15 = 2.1 pm/V

N2; 12/02/2001
T_melt: 161 deg. C
T_NF: 109 deg. C
d33 = 27.8; d31 = 2.0; d15 = 2.3 pm/V

4ECMe
T_melt: 153 deg. C
T_NF: 113 deg. C
d33 = 27.8; d31 = 2.6; d15 = 2.9 pm/V

07/01/2001; 22; IM1; KM1
T_melt: 167 deg. C
T_NF: 104 deg. C
d33 = 27.6; d31 = 2.1; d15 = 2.3 pm/V

NT3F.4
T_melt: 124 deg. C
T_NF: 97 deg. C
d33 = 27.2; d31 = 1.8; d15 = 2.1 pm/V

EC4F
T_melt: 112 deg. C
T_NF: 153 deg. C
d33 = 27.2; d31 = 1.8; d15 = 2.1 pm/V

Nt1
T_melt: 20 deg. C
T_NF: 41.5 deg. C
d33 = 27.1; d31 = 2.7; d15 = 3.4 pm/V

CJG-1-3; 2ECMe; AM1
T_melt: 157.3 deg. C
T_NF: 125.0 deg. C
d33 = 26.9; d31 = 1.7; d15 = 2.0 pm/V

4ECMeF
T_melt: 139 deg. C
T_NF: 109 deg. C
d33 = 26.6; d31 = 2.0; d15 = 2.3 pm/V

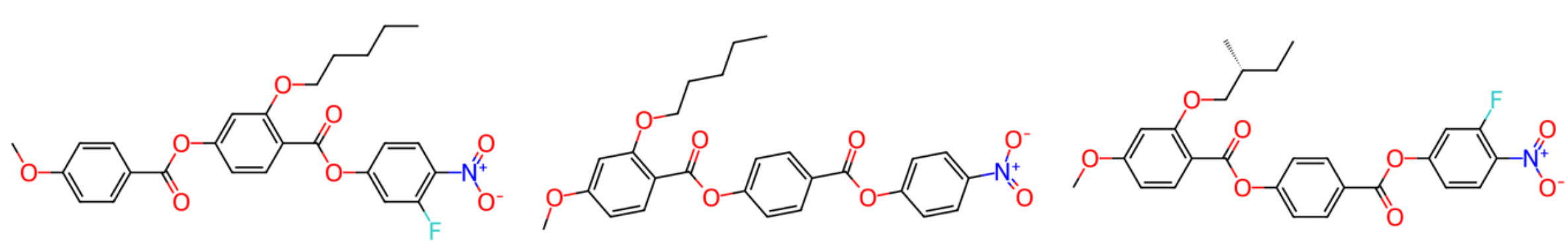

NT3F.5
T_melt: 104 deg. C
T_NF: 70 deg. C
d33 = 26.4; d31 = 1.7; d15 = 2.0 pm/V

M5
T_melt: 132 deg. C
T_NF: 54 deg. C
d33 = 26.4; d31 = 1.8; d15 = 2.1 pm/V

II-4
T_melt: 111 deg. C
T_NF: 54.9 deg. C
d33 = 25.9; d31 = 1.6; d15 = 1.9 pm/V

**Figure SI-7:** Continued.

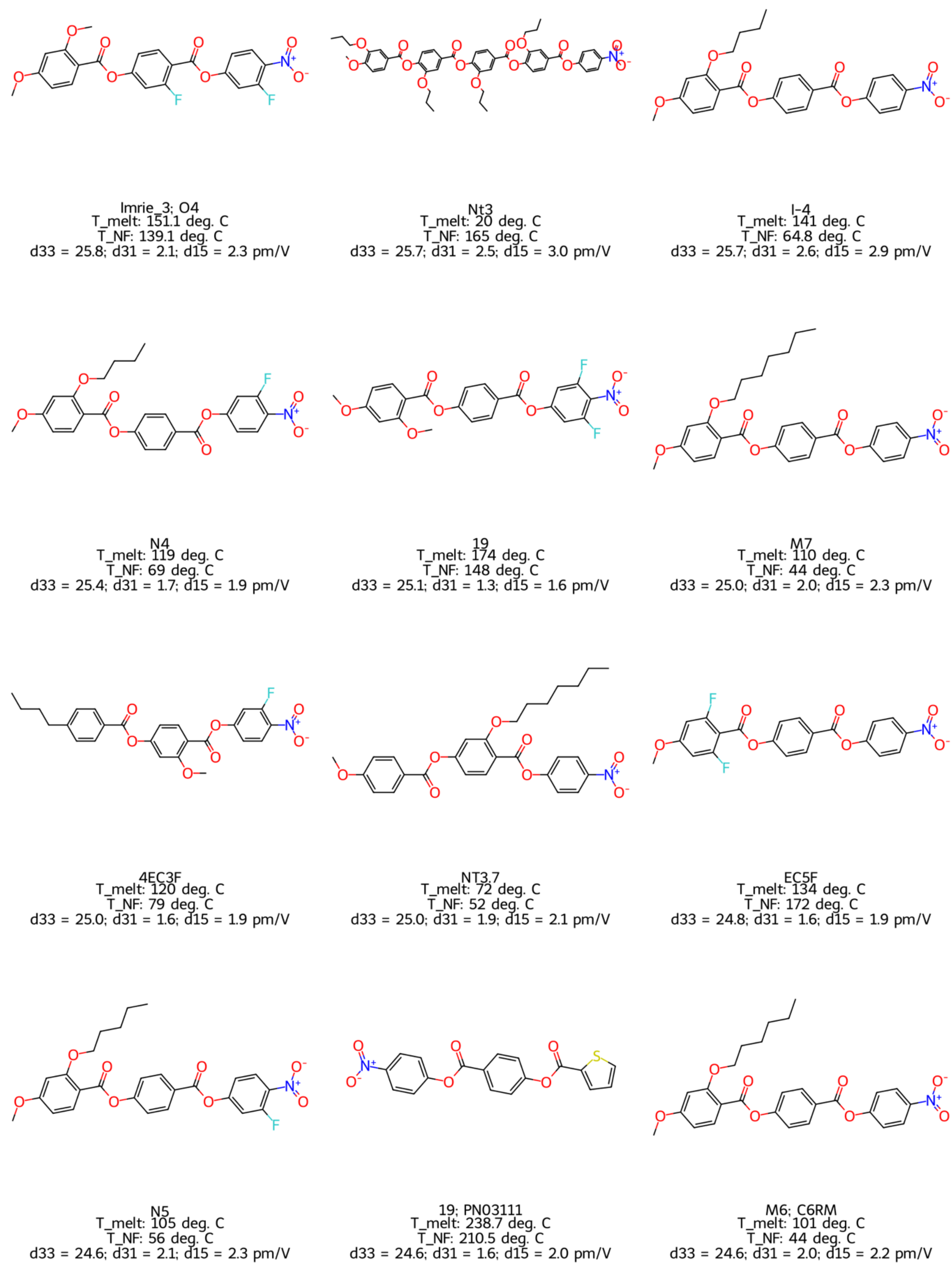

**Figure SI-7:** Continued.

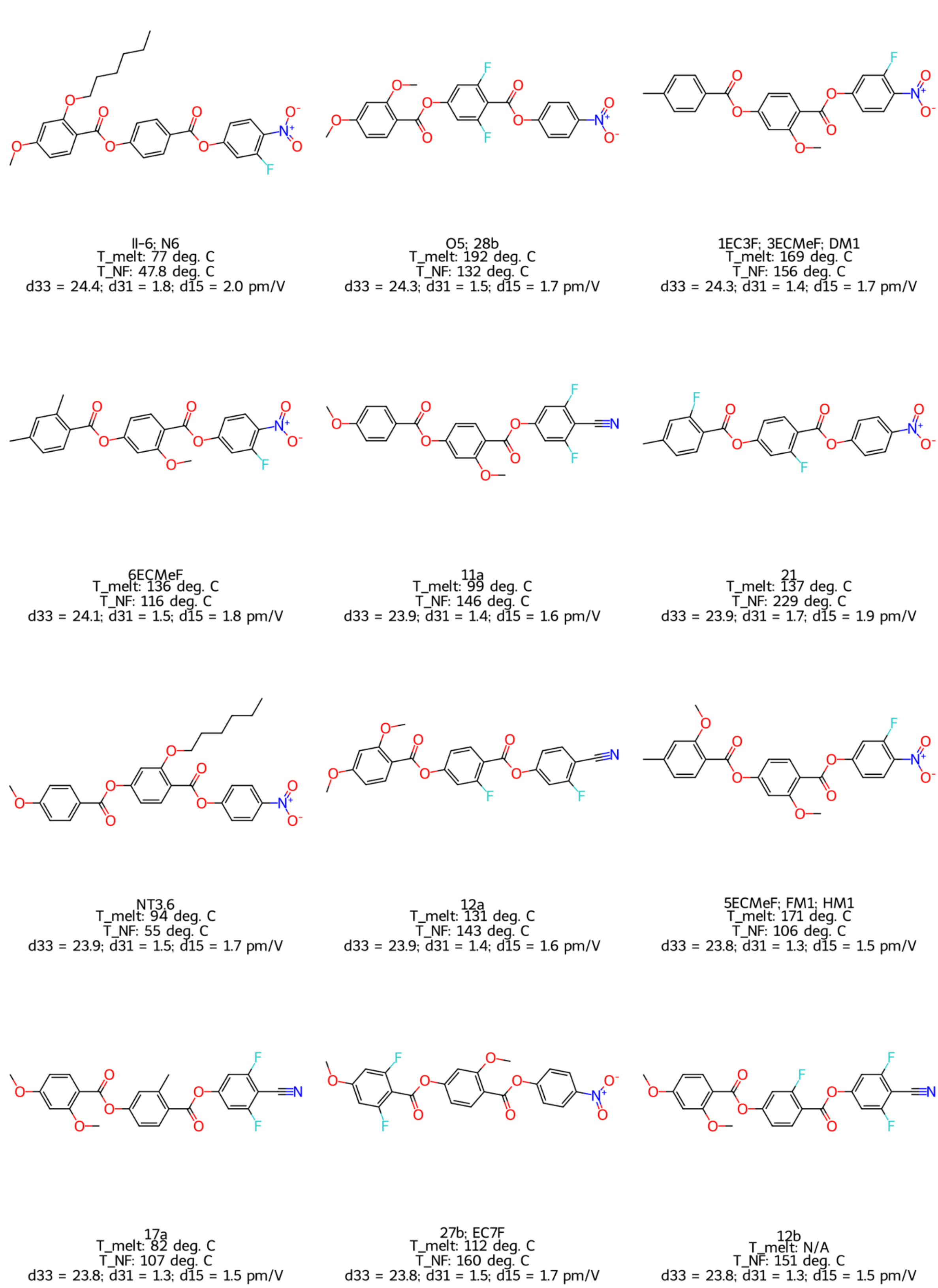

**Figure SI-7:** Continued.

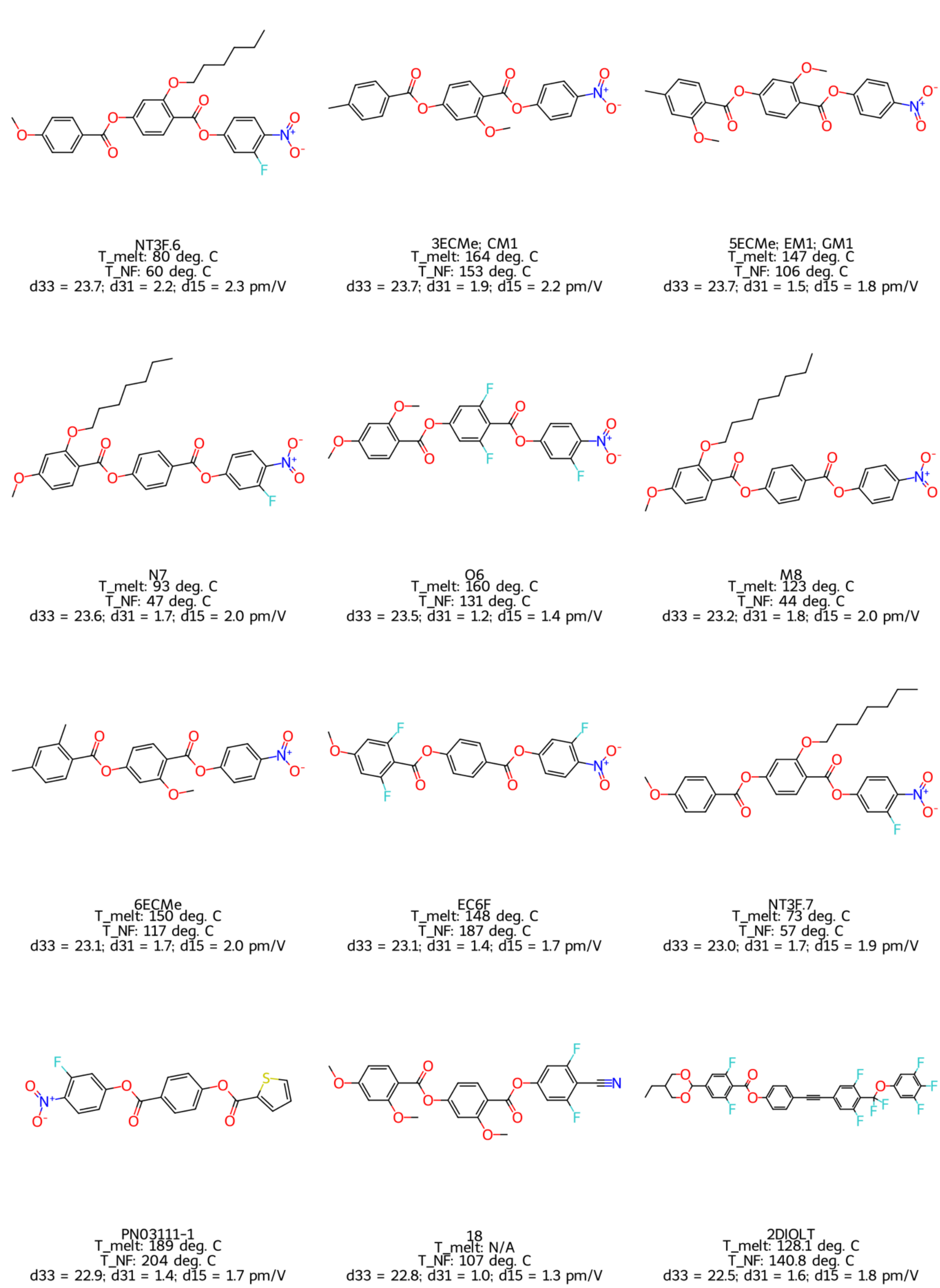

**Figure SI-7:** Continued.

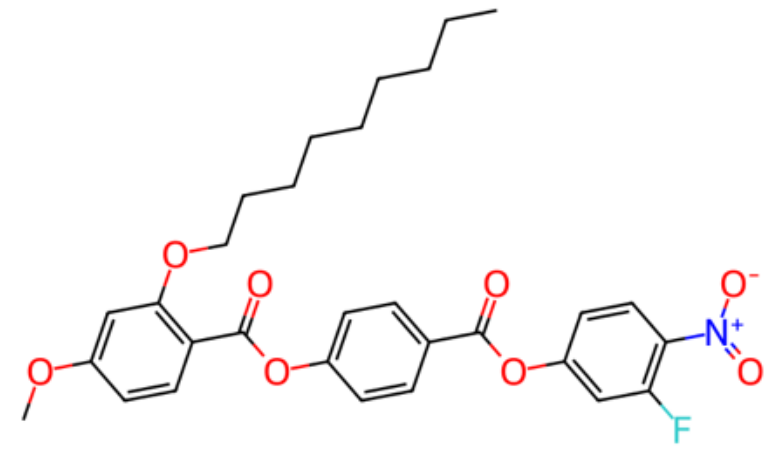

N9
T_melt: 105 deg. C
T_NF: 46 deg. C
d33 = 22.3; d31 = 1.3; d15 = 1.5 pm/V

1DIOLT
T_melt: 124.9 deg. C
T_NF: 178.9 deg. C
d33 = 22.2; d31 = 1.7; d15 = 1.8 pm/V

3DIOLT
T_melt: 95.2 deg. C
T_NF: 127.1 deg. C
d33 = 22.1; d31 = 1.5; d15 = 1.7 pm/V

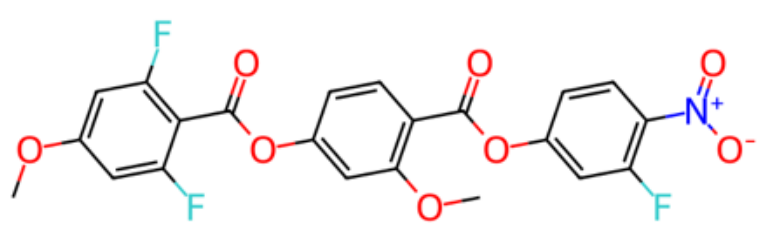

EC8F
T_melt: 140 deg. C
T_NF: 149 deg. C
d33 = 22.1; d31 = 1.3; d15 = 1.5 pm/V

2EC3F
T_melt: 146 deg. C
T_NF: 132 deg. C
d33 = 22.0; d31 = 1.7; d15 = 1.9 pm/V

CJG-1-2; 13; oO1
T_melt: 136.4 deg. C
T_NF: 96.3 deg. C
d33 = 21.8; d31 = 1.4; d15 = 1.7 pm/V

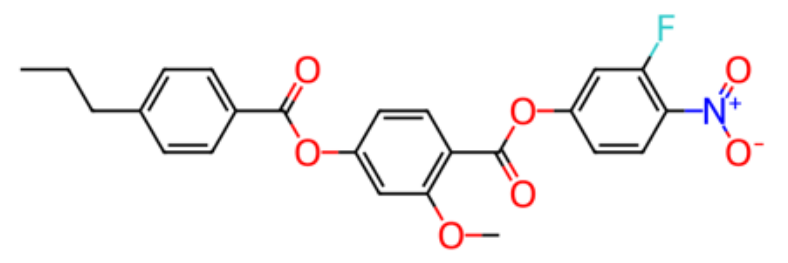

3EC3F
T_melt: 150 deg. C
T_NF: 116 deg. C
d33 = 21.6; d31 = 1.7; d15 = 1.9 pm/V

CJG457
T_melt: 118.5 deg. C
T_NF: 78.1 deg. C
d33 = 21.5; d31 = 1.3; d15 = 1.4 pm/V

12c
T_melt: N/A
T_NF: 76 deg. C
d33 = 21.4; d31 = 1.1; d15 = 1.3 pm/V

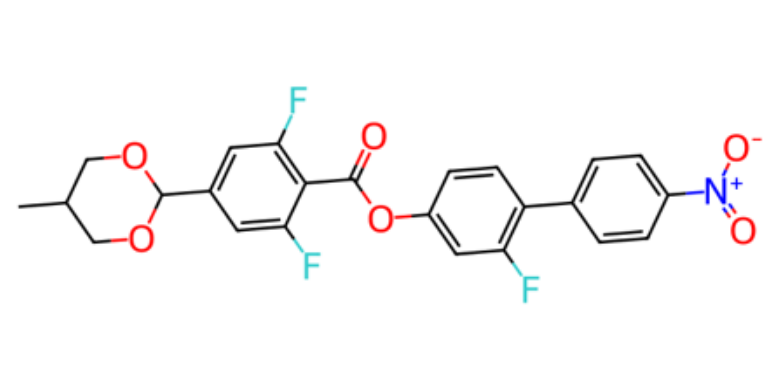

2a
T_melt: 100 deg. C
T_NF: 119 deg. C
d33 = 21.1; d31 = 1.5; d15 = 1.7 pm/V

N8
T_melt: 111 deg. C
T_NF: 48 deg. C
d33 = 20.9; d31 = 1.7; d15 = 1.9 pm/V

coumar_2
T_melt: 203.9 deg. C
T_NF: 120 deg. C
d33 = 20.7; d31 = 1.5; d15 = 1.7 pm/V

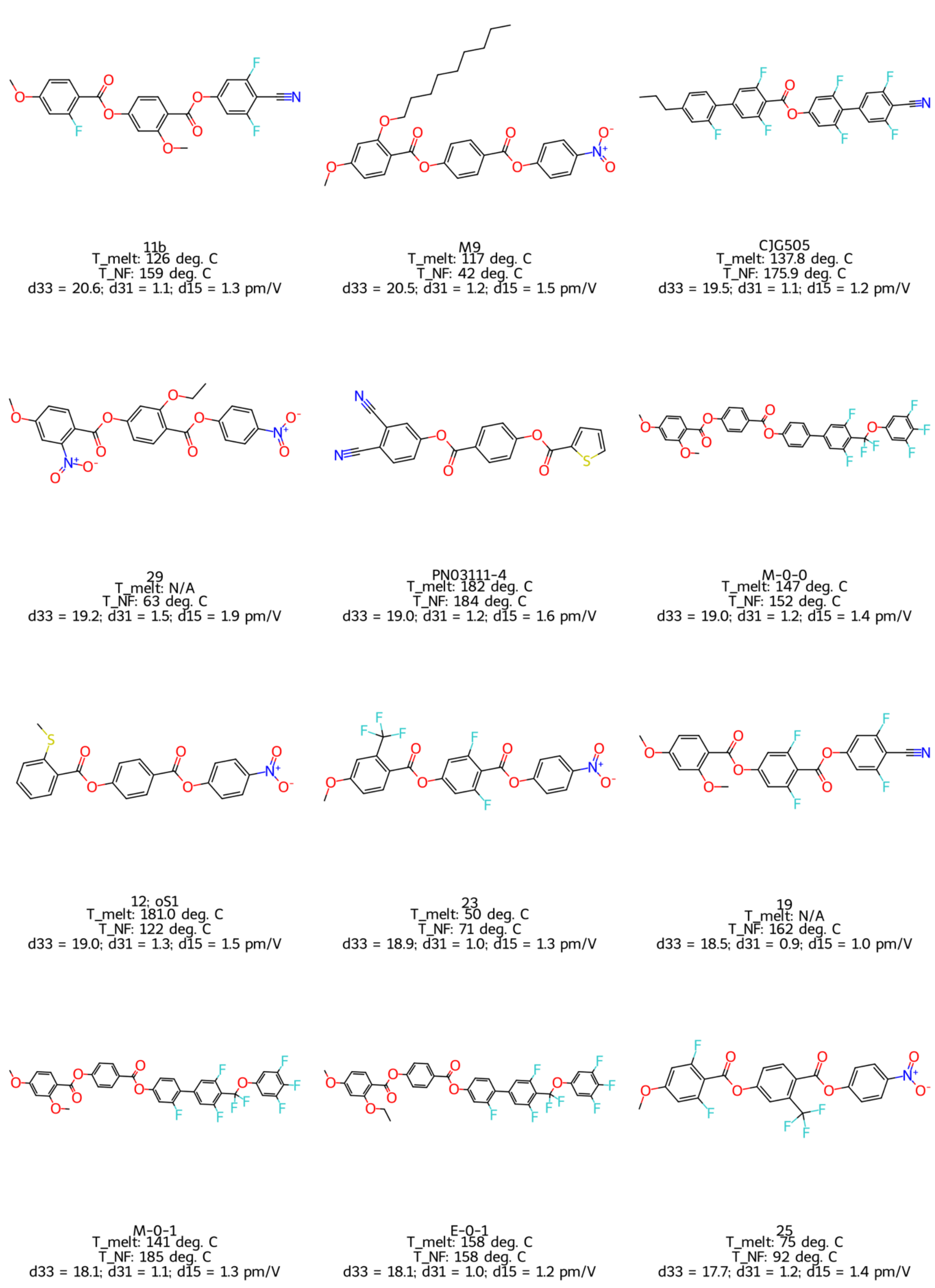

**Figure SI-7:** Continued.

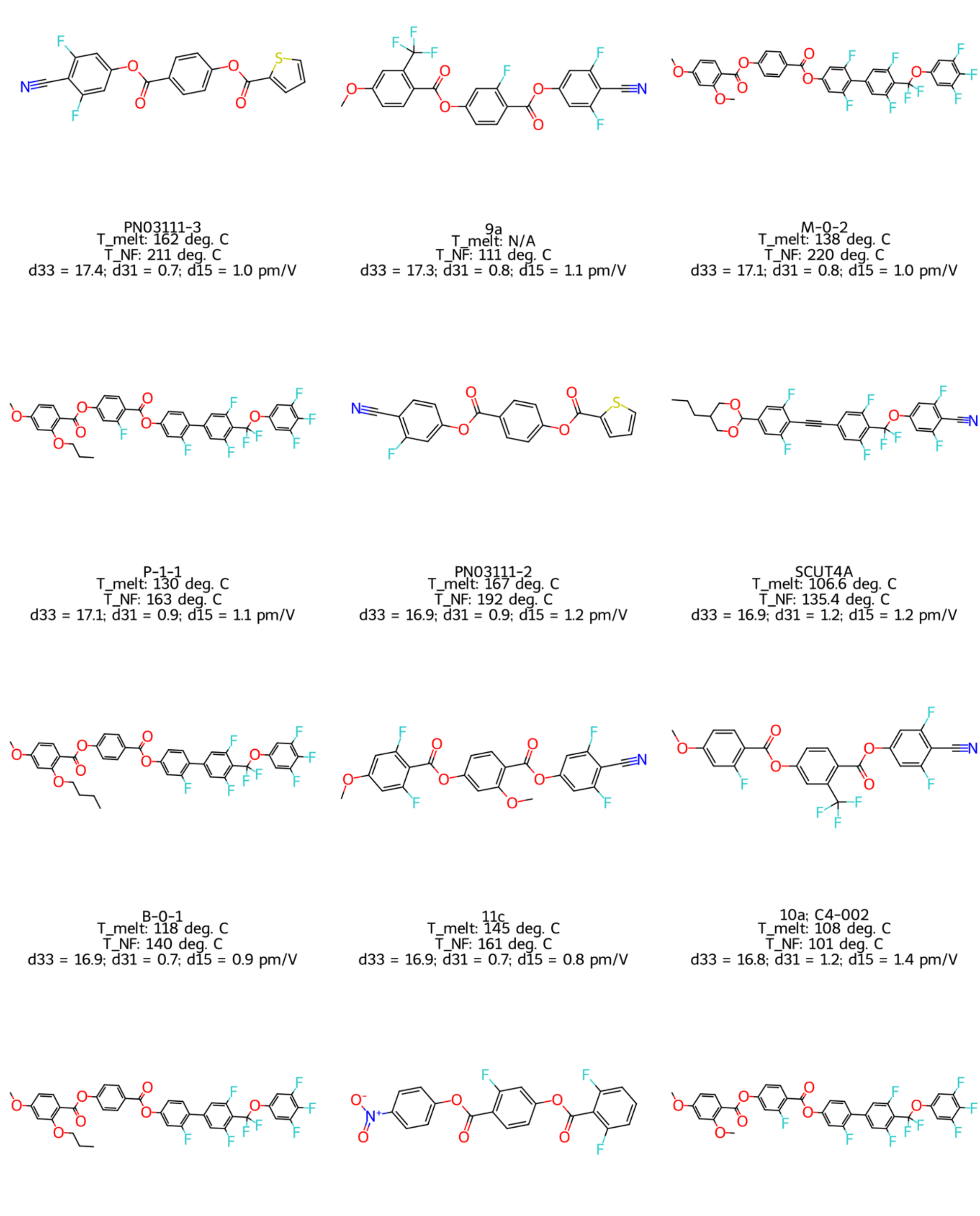

**Figure SI-7:** Continued.

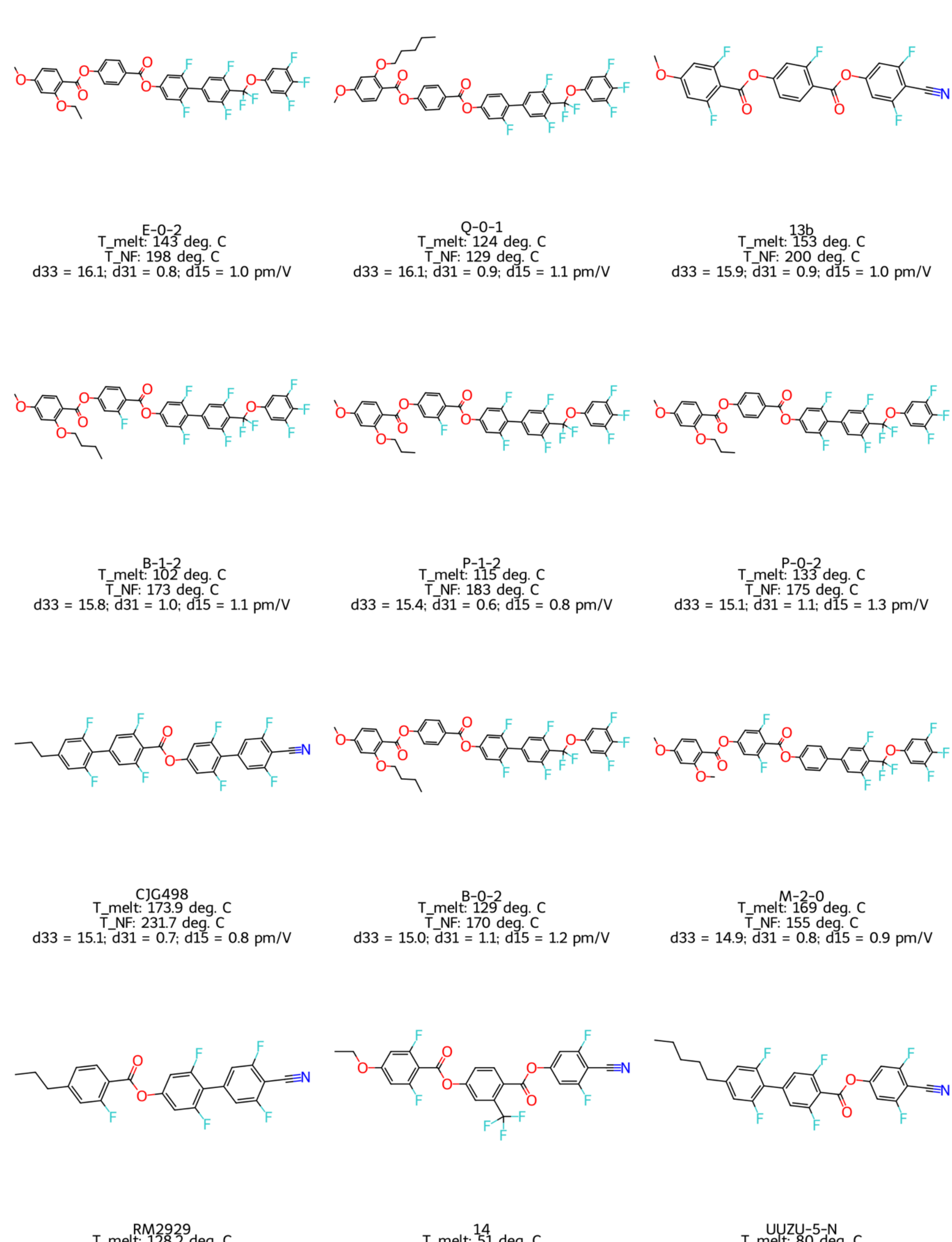

**Figure SI-7:** Continued.

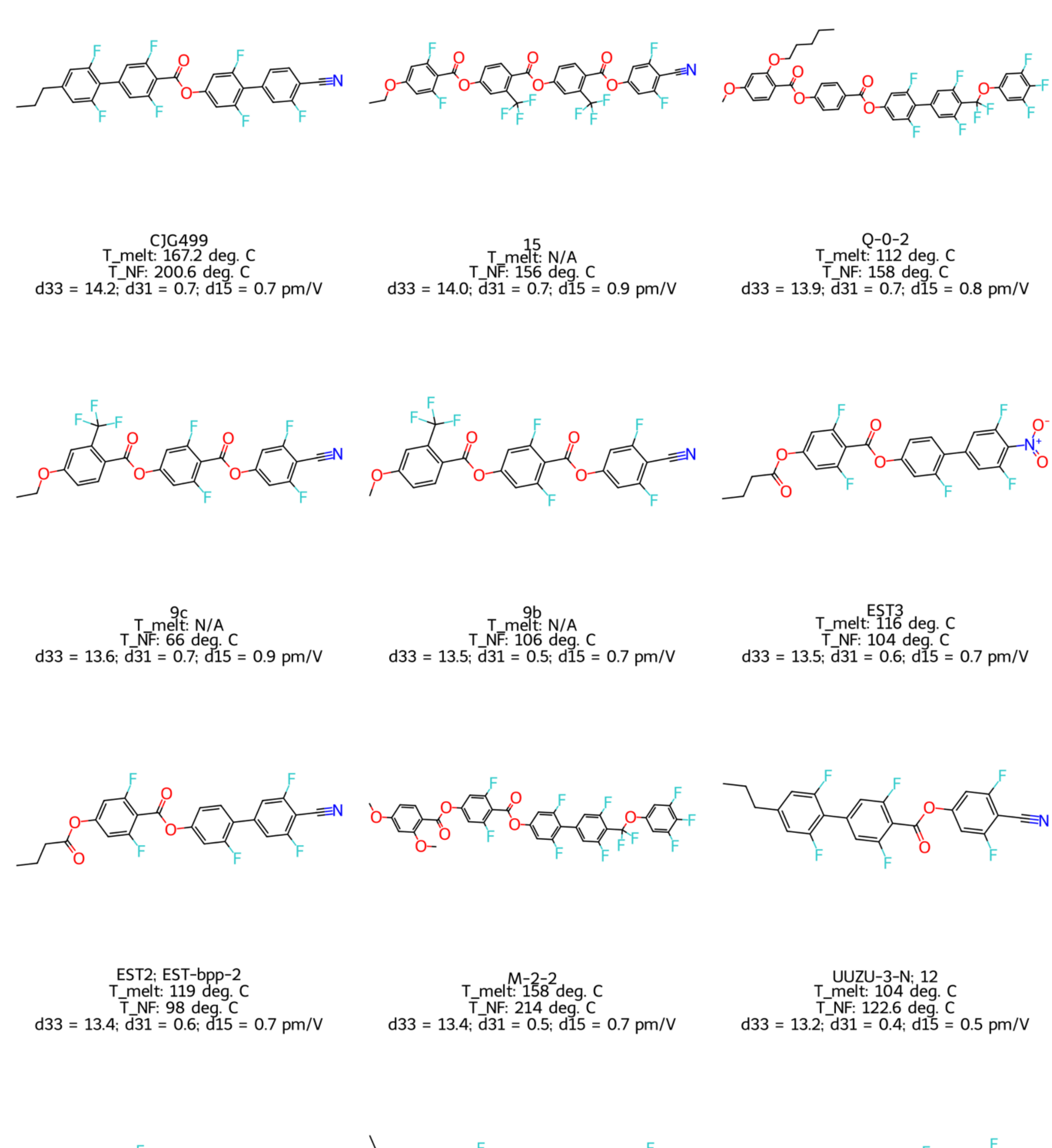

**Figure SI-7:** Continued.

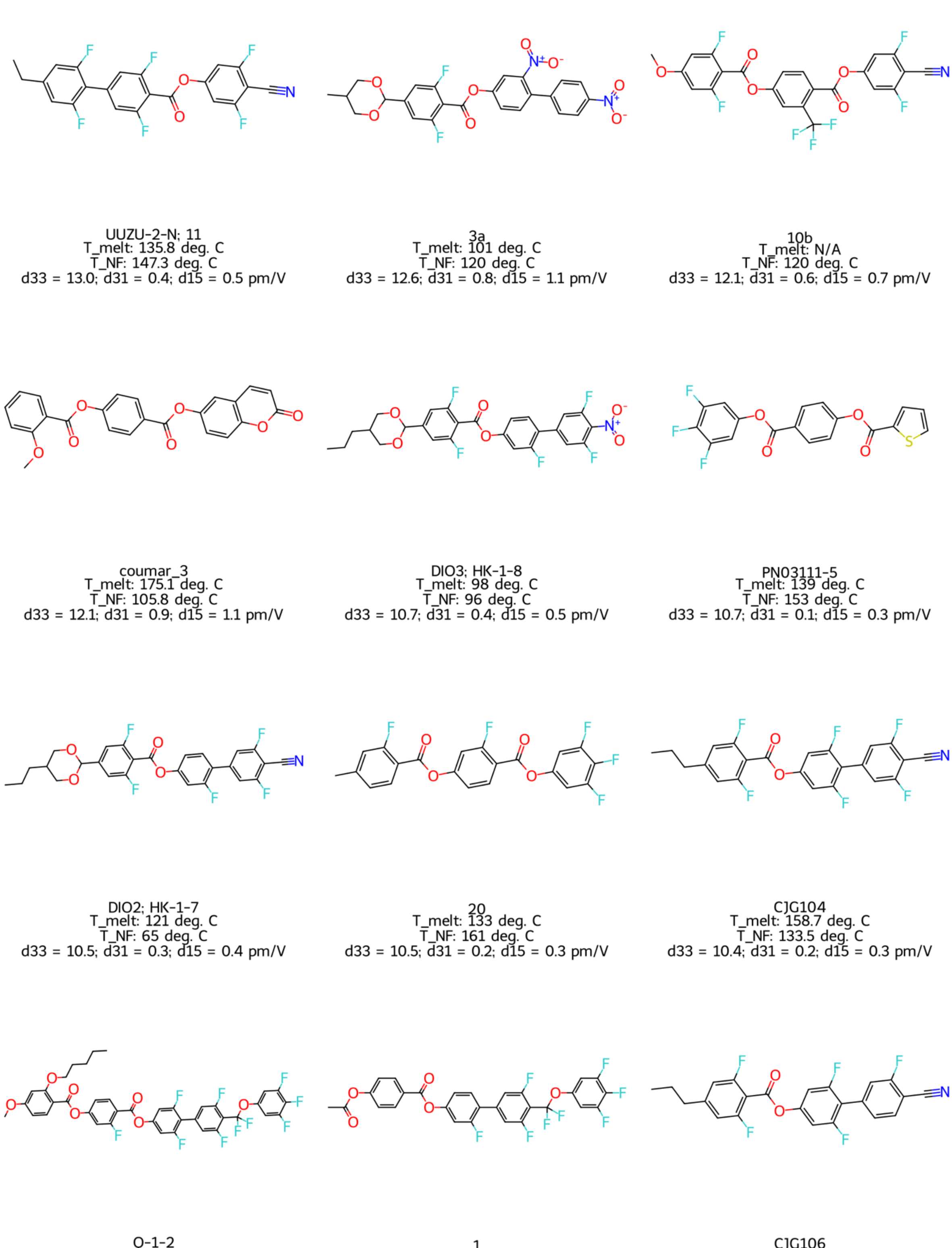

**Figure SI-7:** Continued.

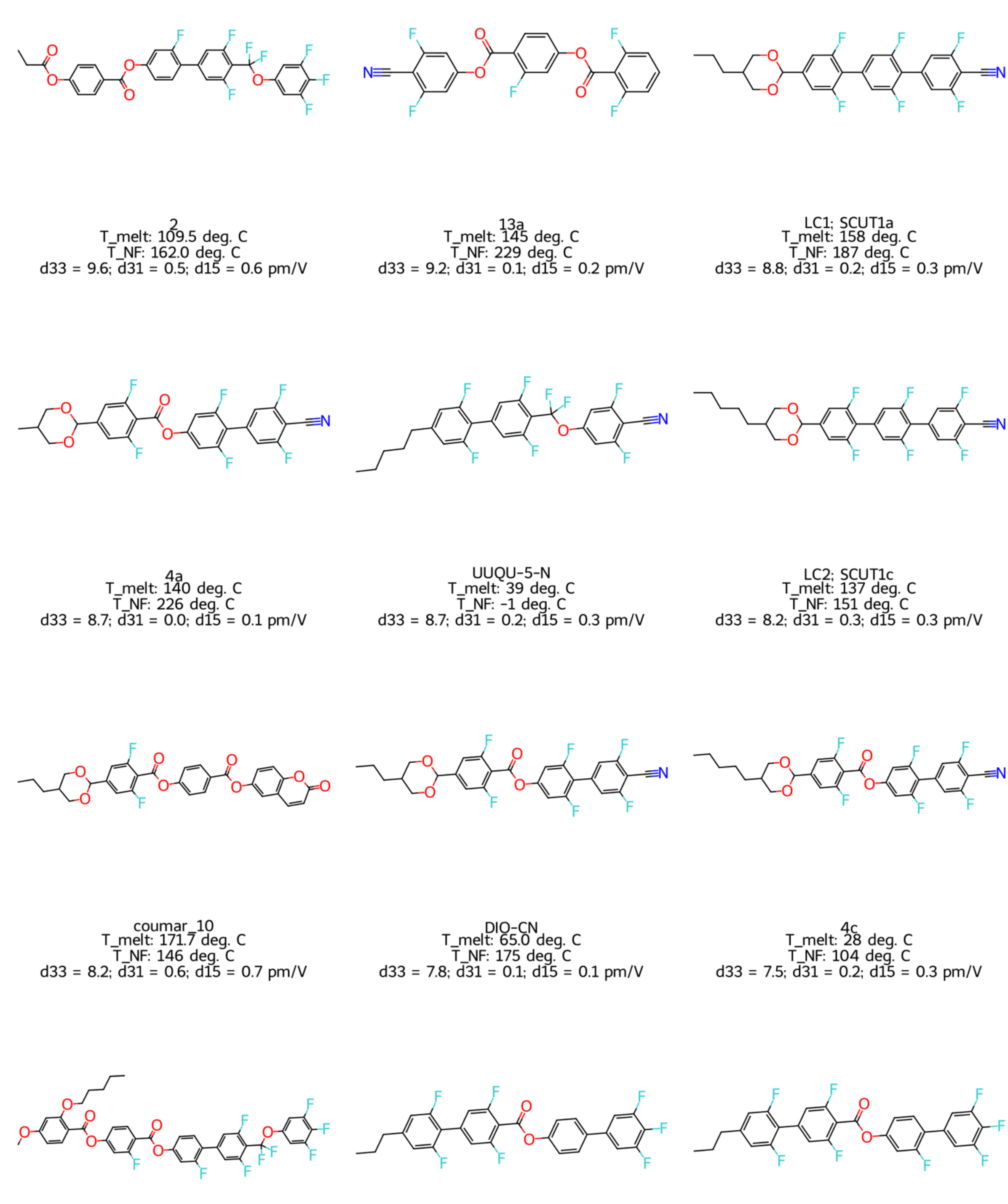

**Figure SI-7:** Continued.

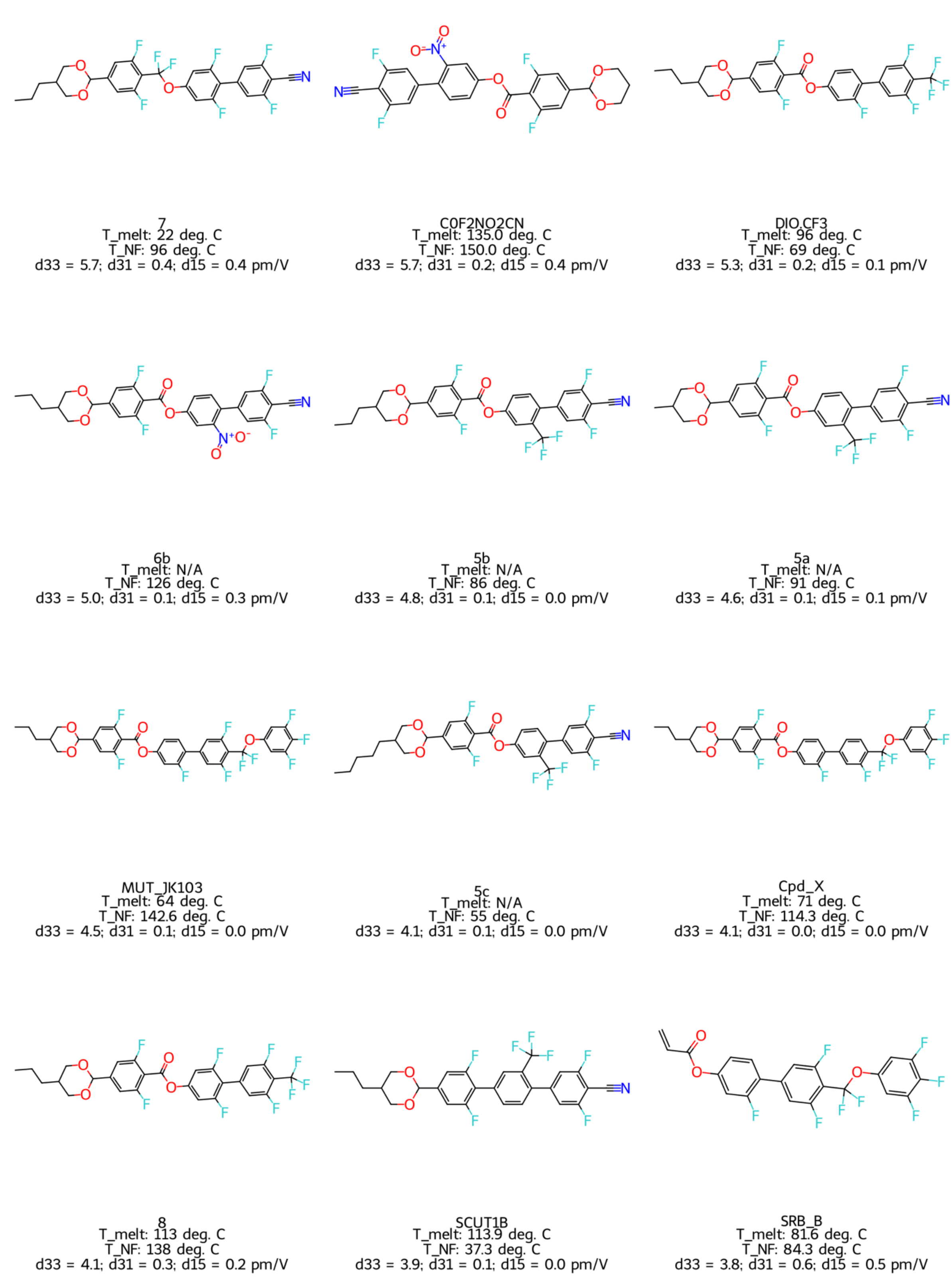

**Figure SI-7:** Continued.

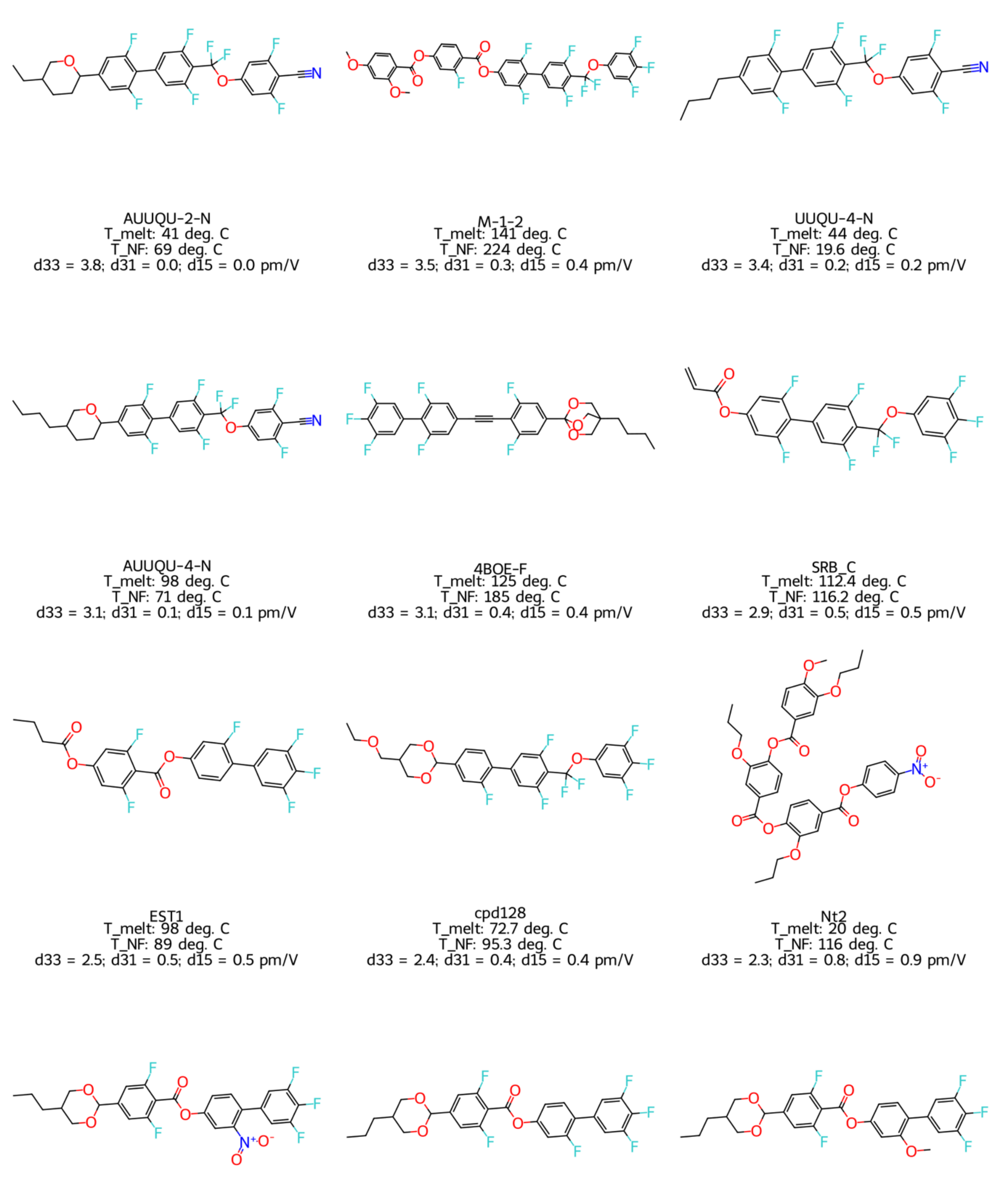

1c
T_melt: N/A
T_NF: 66 deg. C
d33 = 0.6; d31 = 0.4; d15 = 0.3 pm/V

Aya_5B; DIO1; HK-1-1; D3IO-F13
T_melt: 47 deg. C
T_NF: 68.9 deg. C
d33 = 0.6; d31 = 0.7; d15 = 0.7 pm/V

Aya_5C; DIO-F+Ome
T_melt: 43.6 deg. C
T_NF: 66.1 deg. C
d33 = 0.6; d31 = 0.8; d15 = 0.8 pm/V

**Figure SI-7:** Continued.

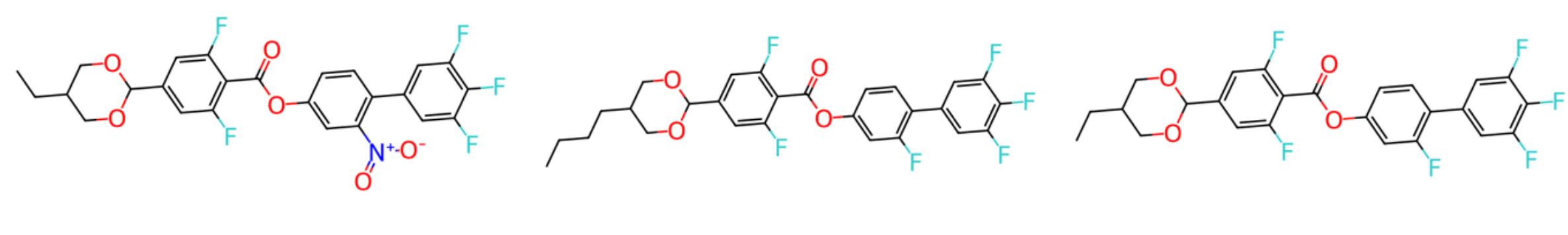

1b
T_melt: 63 deg. C
T_NF: 65 deg. C
d33 = 0.4; d31 = 0.5; d15 = 0.3 pm/V

D4IO-F13
T_melt: 71 deg. C
T_NF: 45 deg. C
d33 = 0.4; d31 = 0.7; d15 = 0.7 pm/V

D2IO-F13
T_melt: 92 deg. C
T_NF: 97 deg. C
d33 = 0.3; d31 = 0.6; d15 = 0.6 pm/V

D1IO-F13
T_melt: 123 deg. C
T_NF: 137 deg. C
d33 = 0.3; d31 = 0.8; d15 = 0.7 pm/V

1a
T_melt: N/A
T_NF: 87 deg. C
d33 = 0.2; d31 = 0.5; d15 = 0.4 pm/V

D5IO-F23
T_melt: 95 deg. C
T_NF: 84 deg. C
d33 = 0.2; d31 = 0.8; d15 = 0.7 pm/V

D2IO-F23
T_melt: 133 deg. C
T_NF: 140 deg. C
d33 = 0.1; d31 = 0.8; d15 = 0.8 pm/V

Aya_5D; DIO+F; D3IO-F23
T_melt: 96.1 deg. C
T_NF: 123.7 deg. C
d33 = 0.1; d31 = 0.7; d15 = 0.7 pm/V

C0F2NO2F3; 0NO2IO
T_melt: 134.6 deg. C
T_NF: 83.4 deg. C
d33 = 0.1; d31 = 0.5; d15 = 0.4 pm/V

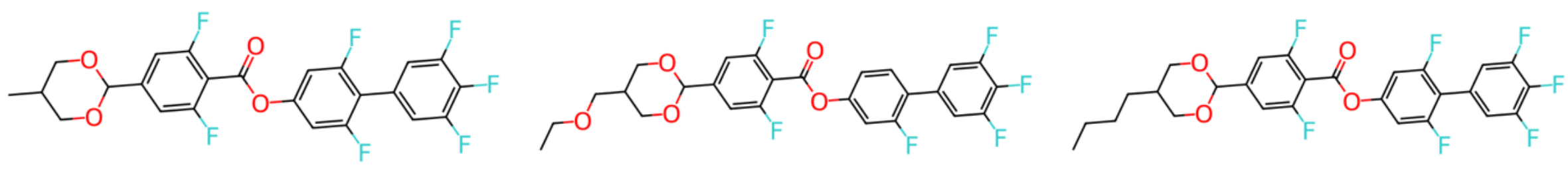

D1IO-F23
T_melt: 157 deg. C
T_NF: 159 deg. C
d33 = 0.1; d31 = 0.8; d15 = 0.8 pm/V

cpd157
T_melt: 92.4 deg. C
T_NF: 135.6 deg. C
d33 = 0.0; d31 = 0.6; d15 = 0.6 pm/V

D4IO-F23
T_melt: 110 deg. C
T_NF: 104 deg. C
d33 = 0.0; d31 = 0.7; d15 = 0.7 pm/V

**Figure SI-7:** Continued.